\documentstyle[pra,eqsecnum,aps]{revtex}

\begin{document}
\draft

\title{Relativistic Collapse Model With Tachyonic Features}
\author{Philip Pearle}
\address{Department of Physics, Hamilton College, Clinton, NY  13323}
\date{\today}

\maketitle

\begin{abstract}

{A finite relativistic model for free particles, which describes the 
collapse of the statevector, is presented. The interaction of particles with a 
classical fluctuating field $w(x)$ causes collapse, as in the (so far) 
physically satisfactory nonrelativistic Continuous Spontaneous Localization (CSL) model.
In previous relativistic models, where $w(x)$ has a white noise spectrum, 
collapse is accompanied by spontaneous creation 
of particles with all energies out of the vacuum 
(amounting to infinite energy/sec-vol). 
In this paper we explore one way 
of eliminating such vacuum excitation. 
We present a formalism which allows an arbitrary spectrum for $w(x)$.  
We point out, in 
lowest order of perturbation theory, 
that it is timelike components of this spectrum which 
are responsible for vacuum excitation  but (appropriate, it seems, to a fundamentally 
nonlocal phenomenon) it is spacelike components which are 
responsible for collapse. We restrict the spectrum of $w(x)$ to that of a 
tachyon of mass $\mu=\hbar /ac\approx 1$eV, where $a\approx 10^{-5}$cm is the 
Ghirardi-Rimini-Weber (GRW) collapse model scale parameter used in CSL.  
However, in higher orders than the first, there is still vacuum excitation 
due to the energy-momentum supplied by particle propagators.   
We opt to explore a simple means of 
eliminating this vacuum excitation: by removing the 
time-ordering operation from the statevector evolution so that 
particle propagators are on the mass-shell. The result is a finite theory:  
there is no vacuum excitation (and no need for renormalization) as 
each vertex describes a finite physical process, the spontaneous emission/absorption 
of a real tachyon by a particle (pair creation/annihilation 
is not permitted by energy-momentum conservation).    
This process may be thought of as analogous to the GRW ``hitting" 
description of collapse.  The cost of no time-ordering is that at each vertex   
a mass $M$ particle's wavefunction can spread by as 
much as $\approx cT(\mu /M)$ in time $T$. This has the result that, 
beginning at 
perturbation order $2M/\mu$, there is a   
nonvanishing probability for the particle to 
be found outside its lightcone: 
for electrons and nucleons, the probability of this 
spacelike transport is negligibly small. Apart 
from this effect, the density matrix of particles in one region 
evolves independently of particles in another spacelike separated region.
We thus obtain a finite, reasonably sensible, relativistic 
collapse model for free particles.}  

\end{abstract}

\pacs{03.65.Bz,11.10.Lm,11.90.+t, 12.90.+b}

\section{Introduction}

\subsection{CSL}\label{sec:CSLintro}

	There is something missing in the foundations of standard quantum theory (SQT) 	
which may motivate one to look for a possible deeper theory\cite{probs}. 
What SQT lacks is the specification
 of necessary conditions for when
an event occurs.  A sufficient condition {\it is} given: 
``upon completion of a measurement by an apparatus," after which one is supposed to 
``collapse the statevector" by hand to a state consistent 
with the outcome of the event.  But this   
sufficient condition is ill-defined: what's a measurement? 
what's an apparatus? when is completion? do events occur in other circumstances?  

	In collapse models, this ill-defined sufficiency condition or ``collapse postulate" 
of SQT is replaced by a well-defined evolution 
of the statevector.   Collapse takes place dynamically, to a state 
consistent with the reality we observe around us, with a probability 
consistent with the predictions of SQT for most (but not all) experiments. 
This 30 year old program\cite{BohmBub,Pearle} 
was reinvigorated by the Spontaneous Localization (SL) model of Ghirardi, Rimini and 
Weber\cite{GRW,Bell} (GRW) a decade ago.  At present, the 
nonrelativistic Continuous Spontaneous Localization (CSL) model\cite{PearleCSL,GPR} 
is the most satisfactory representative of this program.    
  
  The constraints on such a model are very severe.  It must give the results of 
SQT for {\it all} experiments to date. It must generate a rapid collapse for  
a statevector which is a superposition of states of a macro-system (e.g., an apparatus) and    
the result of such a collapse must be a state which we actually observe in Nature.  
However, there should be 
negligible collapse of a statevector which is a 
superposition of states of a micro-system (e.g., of 
a particle undergoing interference).

	CSL consists of two equations. One is the Evolution Equation for the statevector. 
To Schrodinger's equation is added an extra term. It depends upon a 
classical fluctuating field $w({\bf x},t)$ which interacts with 
the number density of particles to cause collapse. 
This term contains two parameters introduced by GRW: $\lambda ^{-1}\approx 10^{16}$sec $\approx 300$ 
million years which characterizes the collapse time of an isolated nucleon 
and $a\approx 10^{-5}$cm which characterizes the distance beyond which collapse is most effective.
It also contains dimensionless parameters to characterize the coupling of $w({\bf x},t)$ to 
different particle types, e.g., nucleon and electron\cite{PearleSquires}.  The second 
equation of CSL is a Probability Rule. It gives the probability that a particular $w({\bf x},t)$ is 
to be found in Nature.  

	Since the evolution of the statevector is different in CSL than in SQT, this allows for 
experimental tests. For example, a possible test is 
observation of a two slit interference pattern of a ``large object'' such as a
$100^{\circ}$A sphere of mercury\cite{Clauser}, since CSL 
predicts that the amplitudes of spatially separated packets 
will change with time\cite{Neutron} whereas SQT does not. While this has 
so far been too difficult to perform, another kind of test {\it has} 
been performed.  

	In CSL, collapse narrows wavefunctions,  
thereby increasing the energy of particles.  This causes ``spontaneous" excitation of 
bound states which then radiate\cite{GR,Squires}.  
Experimental data on ``spontaneous" radiation of X-rays in Ge,
while not yet ruling out CSL or SQT, 
has recently been employed to provide 
valuable constraints on the parameters of CSL\cite{Collett,PRCA}.  It suggests that 
the coupling of $w({\bf x},t)$ to particles may be proportional to their mass, 
supporting the idea of a possible connection between collapse and gravity\cite{GenRel,Diosi}.   
Also recently, it has been 
pointed out that experiments which place a limit on the lifetime of the proton 
could provide an even more stringent test\cite{PearleSquires}, since the collapse mechanism ought to 
excite the quarks in the proton, and the decay products of the resulting unstable particle 
should be observed in proton lifetime experiments.

\subsection{Problems Of The First Relativistic Collapse Model}\label{sec:RCSLintro}

	Since quarks move highly relativistically, 
one must have a relativistic version of CSL---which we shall call RCSL---
in order to make a prediction of the effect of collapse on the proton lifetime.    

	Of course, this is not the only reason to construct a viable relativistic collapse model. 
After all, the relativistic symmetry exists in nature.  John Bell was so concerned about 
the problem of making such a model that his first paper on collapse models\cite{Bell}
and his last paper\cite{Belllastpaper} dealt with it. Collapse 
is a nonlocal behavior and it is not obvious how to make a nonlocal relativistic model, 
but in his first paper he argued that the nonrelativistic GRW model's 
mechanisms did not preclude a relativistic generalization. 

	Indeed, the subsequent discovery of CSL made it 
possible to construct an RCSL model (RCSL$_1$)\cite{PearleRel,GGP,PearleRel2,Dowrick}.   
But, in addition 
to collapsing the statevector, RCSL$_1$ also 
has the undesirable feature of creating infinite energy/sec-vol (i.e., particles of 
all energies) out of the vacuum. In his last paper, Bell 
suggested that this difficulty indicates that
RCSL$_1$ is not nonlocal enough, and he speculated that a resolution of the problem  
might eventually lead to something of fundamental significance.

	The reason for this vacuum excitation is as follows.  
In RCSL$_1$, a scalar field (of mass $\mu = a^{-1}\approx 1$eV) interacts as usual with 
the fermion field, and ``dresses" each localized fermion with an average Yukawa field amplitude. 
Now, $w({\bf x},t)$ also interacts with this scalar field  
in such a way that the dynamics sees the statevector as a 
superposition of different scalar field amplitude states and attempts 
to collapse the statevector to one of these. When the statevector is a superposition 
of different localized fermion states, it is ipso facto a superposition of 
different scalar field amplitude "dresses" and the statevector is thereby  
collapsed toward one such state: this is the desired behavior.  

	However, the vacuum 
is also a superposition of different scalar 
field amplitude states. RCSL$_1$ dynamics 
tries to collapse the vacuum state to one of these. Such 
an alteration of the vacuum state amounts to exciting 
scalar particles out of the vacuum. The result is that the expectation value of the energy 
in each momentum mode of the scalar field increases linearly 
with time at the same constant slow rate of about 1 eV per 300 million years.   
However, there are an infinite number of modes so there is infinite energy production. 
Indeed, in a relativistic theory, if any momentum mode of the vacuum is excited, 
then all must be excited.  
This means that a viable RCSL {\it must have no vacuum excitation whatsoever}.

\subsection{Outline Of Paper}\label{sec:outlineintro}

	The purpose of this paper is to show one way of eliminating vacuum excitation and obtaining
a finite RCSL model.

	In RCSL$_1$, $w({\bf x},t)$ has the statistics of white 
noise, so that its spectrum possesses all four-momenta in equal amounts. 
Although I introduced the idea of describing wavefunction collapse by a 
stochastic differential Schrodinger equation dependent upon white noise\cite{Pearle},  
this was largely for mathematical simplicity.  White noise has all frequencies and physical 
processes do not, so I have always expected that eventually the spectrum of the  
noise would be modified.  However, until this problem of 
vacuum excitation arose there had been no reason to make such a modification.   
A few years ago I proposed a formalism   
which generalizes CSL so that the spectrum of $w({\bf x},t)$ is adjustable\cite{Noise}.  
The price paid is that the statevector's evolution is no longer Markovian
(i.e., its future depends upon a past interval).  Here we shall see that 
a relativistic version of this formalism allows removal of the culpable  
four-momenta which excite particles out of the vacuum. The result is 
a finite relativistic collapse model for free particles.       

	Section \ref{sec:CSLFORMALISM} contains a brief introduction 
to the formalism of CSL, and 
section \ref{sec:CSLexample} gives an example of its use. 
In section \ref{sec:formalism}, 
the generalized formalism which allows $w({\bf x},t)$ to have a colored 
spectrum is introduced. In section \ref{sec:vac}, RCSL$_1$ and its generalized 
version are discussed and it is shown 
that the timelike part of the noise spectrum is responsible 
for producing particles out of the vacuum in lowest order of 
the perturbation theory expansion of the density matrix.

In section \ref{sec:RCSL2}, this is cured 
by making the choice of a spacelike spectrum, that of a 
tachyon of mass $\mu$ ($k^{2}\equiv k_{0}^{2}-{\bf k}^{2}=-\mu ^{2}$). It is shown that the
nonrelativistic ($c\rightarrow \infty$) limit is CSL.  It is also pointed out that 
the scalar field of RCSL$_1$ may be dispensed 
with without altering the physical content of the model  
if the theory just has $w({\bf x},t)$ directly interacting with the particle number density. 
At this point we have a model (which we call RCSL$_{2}$) with no first order 
particle production from the vacuum, with sensible first order 
relativistic collapse behavior and with a sensible nonrelativistic limit.  
  
	However, in higher orders than the first there are still particles produced 
out of the vacuum. This is because these Feynman diagrams contain 
internal particle lines: the associated 
Feynman propagator can convey arbitrary four-momenta from a tachyon at a vertex  
to create particles.  One way to eliminate this 
is to change the evolution equation by relativistically smearing the particle 
number density in such a way that
the four-momenta each vertex allows propagators to exchange with a tachyon is limited. 
We shall purse that approach elsewhere, 
but here (in sections \ref{sec:densmatrix}, \ref{sec:RCSL} and the 
remainder of this paper) we explore a simpler (in the sense of not introducing 
any new functions or constants) but perhaps more radical way of 
eliminating this vacuum excitation: by removing the time-ordering operation from the 
evolution operator. This puts the propagator on the mass shell, thereby 
restricting its four-momenta
and providing a completely finite collapse model 
(RCSL) to all orders. At each vertex, a real particle emits or absorbs a real tachyon.  
Pair production or annihilation is forbidden at each vertex by energy-momentum conservation, 
since a pair's timelike four-momentum cannot equal a tachyon's spacelike four-momentum.  
With no pair production, there is no particle production from the vacuum.  

	A big bonus is that there is no need for renormalization.  
All diagrams are finite as the 
physical process at each vertex is finite.  

	Section \ref{sec:Notime} discusses no time-ordering, pointing out that a perturbation term 
may be thought of as involving successive forward and backwards-in-time 
evolutions (forward with interaction, backwards with free motion) of the statevector, 
which can lead to spacelike transport of particles. Section \ref{sec:ParticleMotion} 
shows how, in time T, the motion of a wavepacket 
in second order of perturbation theory can result in a particle 
of mass M having a wavefunction spread out to $\approx cT(\mu/M)$ resulting, at order 
$2M/\mu$, in a spread to distance $cT$ and, at higher 
orders, in a spread outside the particle's lightcone. However, because of the 
large order and the small phase space volume 
involved, the probability of this is 
extremely small. In section \ref{sec:LocalityRCSL} we show that, with neglect of this effect, 
the model is local in the sense that particles evolve in one region independently of 
particles in a spacelike separated region.

	In section \ref{sec:Results} we give some 
results, the collapse rate 
and energy production rate for a free particle in lowest order.
 	  
	It seems satisfying that, for the fundamentally nonlocal phenomenon of collapse, 
in order to eliminate vacuum excitation in relativistic models, we have been led to 
consider the fundamentally nonlocal relativistic entity: the tachyon. 

	In what follows, 
I have tried to make the body of the paper as compact 
and informative as possible, by putting the main argument in the text, and 
supportive calculations and discussion in appendices.
 
\section{CSL}\label{sec:CSLFORMALISM}

There are two major ideas behind CSL. One is the 
Gambler's Ruin mechanism for ensuring that the 
collapsed states occur with the probabilities of SQT\cite{Pearle}. The other is 
the GRW hitting mechanism for ensuring that the collapse rate is rapid 
for large objects and slow for small ones, and that the collapsed states are what we 
see around us, namely localized objects\cite{GRW}.  

	Consider the collapse dynamics of statevector $|\psi (t)>=a_{1}(t)|1>+a_{2}(t)|2>$  
and, in particular, the  behavior of the squared amplitudes $|a_{i}(t)|^{2}$.  This is 
analogous to the behavior of the dollar amounts $d_{i}(t)$ in the Gambler's Ruin game, 
which is as follows.   
Suppose two gamblers start with $d_{1}(0)$, $d_{2}(0)$ dollars respectively, 
where $d_{1}(0)+d_{2}(0)=100$, and they repeatedly toss a coin 
(analogous to the fluctuating field $w$).  
Depending on the outcome of the toss, one gives a dollar to the other, so the 
$d_{i}(t)$ fluctuate.  Eventually one gambler wins, e.g., $d_{1}(t)/100\rightarrow 1$ and 
$d_{2}(t)\rightarrow 0$ and the game stops. 
It is easy to show that e.g., gambler 1 wins with probability $d_{1}(0)/100$.  With the 
correspondence $d_{i}(t)/100\rightarrow |a_{i}(t)|^{2}$, this means that the amplitudes 
$|a_{i}(t)|^{2}$ fluctuate, and eventually collapse occurs, 
e.g.,  $|\psi (t)>\rightarrow 1|1>+0|2>$ 
with probability $|a_{1}(0)|^{2}$: just the prediction of SQT!

	In GRW's SL model\cite{GRW,Bell} it is supposed that Nature 
subjects the wavefunction of a 
collection of particles not only to the usual Hamiltonian evolution, but also to 
random sudden alterations (``hits"). A hit is the multiplication of 
the wavefunction by a gaussian function $\exp-({\bf x}_{n}-{\bf z})^{2}/2a^{2}$, so that 
the part of the wavefunction which depends 
upon the nth particle's position ${\bf x}_{n}$
is suddenly narrowed to 
width $a$.  The center of the gaussian, ${\bf z}$, is 
randomly chosen according to a certain Probability Rule which depends upon the 
wavefunction, and whose effect is to make it most likely that ${\bf z}$ is located where 
the wavefunction is largest. Such a hit occurs very infrequently on any one particle, 
at the rate $\lambda$.  However, each particle is 
equally likely to be hit, and the wavefunction of particles belonging to an 
apparatus in a superposed state has particles correlated in such a way that 
{\it one} hit on {\it one} apparatus particle effects a collapse 
for the {\it whole} N--particle apparatus, so that such a collapse occurs 
rapidly, in time $1/\lambda N$.
	
		In CSL\cite{PearleCSL,GPR} in effect a hit occurs every $\Delta t$ sec, 
except that the hit wavefunction (instead of replacing the original wavefunction) is 
multiplied by an amplitude $\sim\Delta t$ and added to 
the original wavefunction.  In this way the wavefunction shape 
fluctuates, and the gambler's ruin dynamics is obtained.  Also, the GRW hitting process 
destroys the (anti) symmetry of the wavefunction, and that problem is cured in CSL.  

	The CSL modified Schrodinger equation for the statevector $|\psi,t>_w$ 
which evolves under 
the influence of the fluctuating field $w$ is 
	
\begin{equation}{d|\psi,t>_w\over dt}= -iH|\psi,t>_w - 
{1\over 4\lambda}\int d{\bf x}[w({\bf x},t)
-2\lambda A({\bf x})]^2|\psi,t>\label{2.1}\end{equation}

\noindent Remarkably, this is a linear equation.  All previous 
models had nonlinear equations, and it was suspected this had to be the case. It 
should be remarked that it is this linearity which makes relativistic 
generalizations easy.  The 
operator A({\bf x}) is (apart from a factor $\sim a^{-3/2}$) essentially the 
number of particles in a volume $a^{3} \approx 10^{-15}$cm$^{3}$ centered around ${\bf x}$:  

\begin{equation}A({\bf x})\equiv 
	{1\over (\pi a^{2})^{3\over 4}}\int d{\bf z}N({\bf z}) 
	e^{\displaystyle -{({\bf x}-{\bf z})^{2}
	\over 2a^{2}}}\eqnum{2.1a}\label{2.1a}\end{equation}
	
\noindent ($N({\bf z})\equiv \xi^{\dagger}({\bf x})\xi({\bf x})$ is the particle 
number density operator and $\xi({\bf x})$ is 
the particle annihilation operator at ${\bf x}$). 
 
	CSL is completed by giving the probability density functional $P_{T}(w)$ that 
a particular realization of the field 
$w({\bf x},t)$ ($0\leq t\leq T$) occurs:

\begin{equation}P_{T}(w)=\thinspace_{w}\negthinspace\negthinspace
<\psi,T|\psi,T>_w\label{2.2}\end{equation}
	
\noindent The statevector evolution (\ref{2.1}) is nonunitary, so 
the norm of the statevector $|\psi,t>_w$ 
changes with time: Eq. (\ref{2.2}) says that fields $w({\bf x},t)$ which lead to 
statevectors of largest norm are most likely. Conservation 
of probability is $\int Dw P_{T}(w)=1$,  
where the functional integration element is $Dw\sim\prod_{{\bf x},t} dw({\bf x},t)$.  

	The solution of (\ref{2.1}), in  
the ``collapse interaction picture" where the statevector evolution is 
solely due to collapse dynamics, is
	
\begin{equation}|\psi,T>={\cal T}e^{-{1\over 4\lambda}\int_{T_{0}}^{T}dtd{\bf x}
	[w({\bf x},t)-2\lambda A({\bf x},t)]^2}|\psi,T_{0}>\label{2.3}\end{equation}
	
\noindent where ${\cal T}$ is the time ordering operator and 
$A({\bf x},t)\equiv \exp iHtA({\bf x})\exp -iHt$. 
 
	Eq. (\ref{2.3}) may also be written, using (\ref{2.1a}), as 

\begin{equation}|\psi,T>={\cal T}e^{-{1\over 4\lambda}\int_{T_{0}}^{T}dtd{\bf z}d{\bf z}'
	[w'({\bf z},t)-2\lambda N({\bf z},t)]e^{-{1\over 4a^{2}}({\bf z}-{\bf z}')^{2}} 
	[w'({\bf z}',t)-2\lambda N({\bf z}',t)]}|\psi,T_{0}>\label{2.4}\end{equation}
	
\noindent where $w'$ is related to $w$ by $w({\bf x},t)=
\int d{\bf z}w'({\bf z},t)(\pi a^{2})^{- {3\over 4}}\exp -({\bf x}-{\bf z})^{2}/2a^{2}$ 
(we drop the prime hereafter). It is this form which will prove to be most valuable 
in section \ref{sec:formalism} for generalizing the statevector evolution.

\subsection{CSL Example}\label{sec:CSLexample}

	To see how (\ref{2.4}) leads to collapse for individual statevectors, let $H=0$ (so we are 
not concerned with how the usual dynamics competes with the 
collapse dynamics) and consider particles in a superposed state.  
Suppose that the initial statevector is 
$|\psi,T_{0}>=\sum_{i} c_{i}|n_{i}>$, where 
$N({\bf x})|n_{i}>=n_{i}({\bf x})|n_{i}>$, and the $n_{i}({\bf x})$ represent distinctly different 
particle density distributions.  Then Eqs. (\ref{2.4}) and (\ref{2.2}) become, respectively, 
 
\begin{mathletters}
\label{2.5}
\begin{equation}
|\psi,T>=\sum_{i} c_{i}|n_{i}>e^{-{1\over 4\lambda}\int_{T_{0}}^{T}dtd{\bf x}d{\bf x}'
	[w({\bf x},t)-2\lambda n_{i}({\bf x})]e^{-{1\over 4a^{2}}({\bf x}-{\bf x}')^{2}} 
	[w({\bf x}',t)-2\lambda n_{i}({\bf x}')]}\label{2.5a}
	\end{equation}
	
\begin{equation}
P_{T}(w)=	\sum_{i} |c_{i}|^{2}
e^{-{1\over 2\lambda}\int_{T_{0}}^{T}dtd{\bf x}d{\bf x}'
	[w({\bf x},t)-2\lambda n_{i}({\bf x})]e^{-{1\over 4a^{2}}({\bf x}-{\bf x}')^{2}} 
	[w({\bf x}',t)-2\lambda n_{i}({\bf x}')]}\label{2.5b}
	\end{equation}	
\end{mathletters}
	
\noindent As $T$ increases, any $w({\bf x},t)$ which makes one term in the 
sum in Eq. (\ref{2.5a}) 
or (\ref{2.5b}) large makes the other terms small.  For example, suppose that 
$w({\bf x},t)=2\lambda n_{i}({\bf x})$: then, while 
the $i$th exponential $=1$, the $j$th 
exponential in (\ref{2.5a}) has the value 

\begin{equation}e^{-\lambda (T-{T_{0})\int d{\bf x}d{\bf x}'
	[n_{i}({\bf x})- n_{j}({\bf x})]e^{-{1\over 4a^{2}}({\bf x}-{\bf x}')^{2}} 
	[n_{i}({\bf x}')- n_{j}({\bf x}')]}}\label{2.6}\end{equation}
	
\noindent which approaches $0$ for large $T$. This is also the behavior if, more generally, 
$w({\bf x},t)$ almost always fluctuates uniformly about $2\lambda n_{i}({\bf x})$ 
(for \emph{all} other behaviors of $w$, \emph{all} the terms in (\ref{2.5}) 
asymptotically vanish). Thus the terms in Eqs. (\ref{2.5}) evolve toward regions 
of disjoint support in $w$-space   When $w$ lies in the $i$th region, 
$|\psi,T>\rightarrow\sim |n_{i}>$ according to 
(\ref{2.5a}), while the integrated probability 
over the $i$th region $\int DwP(w)\rightarrow |c_{i}|^{2}$ according to (\ref{2.5b}).

	The density matrix which follows from Eqs. (\ref{2.2}),(\ref{2.3}) is 
	
\begin{eqnarray}\rho (T)\equiv&&\int DwP(w){|\psi,T><\psi,T|\over <\psi ,T|\psi ,T>}\nonumber\\	 
	&&={\cal T}e^{-{\lambda\over 2}\int_{T_{0}}^{T}dtd{\bf x}d{\bf x}'
	[N({\bf x},t)\otimes 1-1\otimes N({\bf x},t)]
	e^{-{1\over 4a^{2}}({\bf x}-{\bf x}')^{2}}
	[N({\bf x}',t)\otimes 1-1\otimes N({\bf x}',t)]}\rho(T_{0})\label{2.7}\end{eqnarray}

\noindent We are employing the notation $A\otimes B\rho=A\rho B$, and ${\cal T}$ time-reverses 
operators to the right of $\rho(T_{0})$.  

	The collapse behavior of the previous example 
may easily be seen in the decay of $\rho (T)$'s off-diagonal density matrix 
elements:

\begin{equation}<n_{i}|\rho (T)|n_{j}>=c_{i}c_{j}^{*}
	e^{-{\lambda\over 2}(T-T_{0})\int d{\bf x}d{\bf x}'
	[n_{i}({\bf x})- n_{j}({\bf x})]
	e^{-{1\over 4a^{2}}({\bf x}-{\bf x}')^{2}}
	[n_{i}({\bf x}')- n_{j}({\bf x}')]}\label{2.8}\end{equation}

\noindent	As a simple example, suppose  $|n_{1}>$ and $|n_{2}>$ each describe a clump of 
N particles with dimensions $<<a$, but 
with centers of mass of the two states at a distance $>>a$ apart.  
Then the gaussian in Eq. (\ref{2.8}) $\approx 1$ if ${\bf x}$, ${\bf x}'$ are both located 
in region 1 (or both in 2) and $\approx 0$ if 
${\bf x}$,  ${\bf x}'$ are located in regions 1 and 2 respectively. Therefore, 
Eq. (\ref{2.8}) yields 

\begin{equation}<n_{1}|\rho (T)|n_{2}>=c_{1}c_{2}^{*}
	e^{-\lambda N^{2}(T-T_{0})}\label{2.9}\end{equation}
	
\noindent illustrating how the collapse rate is large for a superposition of states describing a 
large number of particles in different locations.

	It should be emphasized that the dynamics of an individual statevector under a probable 
$w(x)$ allows nonlocal influences whose outcome, however, is not controllable.  
For example, a particle whose wavefunction describes it as in 
an equal superposition of ``here" plus ``there" may be voluntarily entangled with a 
position measuring apparatus located near ``here," by turning the apparatus on to measure whether the 
particle is ``here".  As a result, for half the probable 
$w(x)$'s, the wavepacket ``there" will disappear, and for 
the other half it will grow to total probability 1.  On the other hand, because the 
field $w(x)$ which determines the outcome is not controllable, for the ensemble of statevectors 
evolving under all possible $w(x)$'s, no nonlocal communication is possible.  The proof of this for 
RCSL is given in section \ref{sec:LocalityRCSL}: the comparable proof for 
CSL is essentially the same

\section{Generalized Formalism}\label{sec:formalism}

   We now postulate, as a generalization of the CSL evolution 
Eq. (\ref{2.4}), the statevector evolution
   
\begin{equation}|\psi ,T>={\cal T} e^{-{1\over 4\gamma}\int_{T_{0}}^{T}dxdx'
   			[w(x)-2\gamma F(x)]G(x-x')[w(x')-2\gamma F(x')]}|\psi ,T_{0}>\label{3.1}\end{equation}
   			
\noindent In Eq. (\ref{3.1}), 
$x\equiv (t,{\bf x})$, $dx\equiv dtd{\bf x}$, $F(x)$ is a family of operators which 
commute at equal times, and $G(x-x')$ is a positive definite even function, i.e., its Fourier 
transform $\tilde G (k)$ is real, even and $\geq 0$.  In addition, 
we retain the probability rule (\ref{2.2}). 

	Eq. (\ref{3.1}) reduces to the CSL evolution Eq. (\ref{2.4}) when
		
\begin{mathletters}\label{3.2}	  
\begin{eqnarray}\gamma=&&\lambda\label{3.2a}\\
G(x-x')=&&\delta (t-t')e^{ -({\bf x}-{\bf x}')^2/4a^{2}}\label{3.2b}\\
F({\bf x},T_{0})\equiv &&N({\bf x})\label{3.2c}\\
F(x)=&&e^{iHt}F({\bf x})e^{ -iHt}\label{3.2d} \end{eqnarray}
\end{mathletters}

	If $G(x-x')\sim \delta (t-t')$, as in Eq. (\ref{3.2b}),  
then $w({\bf x},t)$ is uncorrelated at different times and $|\psi,T>$ satisfies a 
stochastic differential equation (e.g. Eq. (\ref{2.1})), but this is not so in general.

	   When $G(x-x')\not\sim \delta (t-t')$, then $w({\bf x},t)$ 
is correlated at different times and the statevector evolution is no longer Markovian. 
In particular, the Markovian identity 
$\int_{T}^{T'}DwP_{T'}(w)=P_{T}(w)$ enjoyed by CSL no longer holds. We take this to mean that 
the probability estimate of $w({\bf x},t)$  
(regarded as the best measure possible based upon limited information, namely 
lack of knowledge of the future values of w\cite{Cox}) improves with time: strictly speaking, 
only $P_{\infty}(w)$ gives the correct probability of $w$. However, 
for $G$ utilized in this paper, $P_{T}(w)$ essentially changes negligibly with increasing T. 

	That this generalized formalism produces collapse for 
individual statevectors is shown in Appendix \ref{app:A}.

	The density matrix which follows from (\ref{3.1}), (\ref{2.2}) is 	
	
\begin{eqnarray}\rho (T)&& \equiv\int DwP_{T}(w){|\psi,T><\psi,T|\over <\psi ,T|\psi ,T>}\nonumber\\	 
	&& = {\cal T}e^{-{\gamma\over 2}\int_{T_{0}}^{T}dxdx'
	[F(x)\otimes 1-1\otimes F(x)]G(x-x')
	[F(x')\otimes 1-1\otimes F(x')]}\rho(T_{0})\label{3.3}\end{eqnarray}

	It is useful and illuminating to write the statevector (\ref{3.1}) and 
density matrix (\ref{3.3}) as Fourier transforms (Appendix \ref{app:B}):

\begin{equation}|\psi ,T>={\cal T}\int D\eta e^
{-\gamma\int_{-\infty}^{\infty} dxdx'\eta(x)G^{-1}(x-x')\eta(x')}
e^{i\int_{T_{0}}^{T}dx\eta(x)[w(x)-2\gamma F(x)]}|\psi ,{T_{0}}>\label{3.4}\end{equation}

\begin{equation}\rho (T)={\cal T}\int D\eta e^
{-2\gamma\int_{-\infty}^{\infty} dxdx'\eta(x)G^{-1}(x-x')\eta(x')}
e^{-i2\gamma\int_{T_{0}}^{T}dx\eta(x)[F(x)\otimes 1-1\otimes F(x)]}\rho ({T_{0})}
\label{3.5}\end{equation}

\noindent This shows that the statevector and density 
matrix may be written as 
superpositions of unitary evolutions  
$\exp-i2\gamma\int_{T_{0}}^{T}dx\eta(x)F(x)$, describing  
the interaction of a classical ``noise" field $\eta (x)$ with the operator $F(x)$. 
(The superposition has a Gaussian weight and, in the case of Eq. (\ref{3.4}), 
also a phase factor weight.) 
Thus it is actually the spectrum of $\eta (x)$ (the inverse of the spectrum of $w(x)$) 
which excites the 
quantum system\cite{Noise}.  
Eq. (\ref{3.5}) illustrates the well known fact that the density matrix arising from an ensemble of 
``true collapse" evolutions (as discussed here) can be equal to a density 
matrix obtained from an ensemble of non-collapse 
unitary evolutions. 

	Eq. (\ref{3.5}) is particularly useful for calculations 
because it enables one to employ Feynman techniques in evaluating the 
perturbation expansion of the density matrix.

\section{RCSL$_{1}$ And Vacuum Excitation}\label{sec:vac}

	We now apply this formalism to a specific relativistic model.  
	
	In Eq. (\ref{3.1}) let 
$F(x)=\phi (x)$ be a scalar field of mass $\mu$, and also let the Hamiltonian $H$ be such that 
$\phi (x)$ ``dresses" a particle of mass $M$.  The ``dress" 
coupling should be 
$g\phi (x):\negthinspace\negthinspace\bar\psi (x)\psi (x)\negthinspace\negthinspace:$, 
where $\psi (x)$ is a Dirac fermion field, but I do not wish to clutter up 
the calculations with Dirac algebra. Therefore I shall use a boson particle $\psi (x)$, 
with coupling 
$g(2M)\phi (x):\negthinspace\negthinspace\psi (x)\psi (x)\negthinspace\negthinspace:$ (the 
factor $(2M)$ keeps $g$ dimensionless). In the usual 
interaction picture (fields obey free-field evolution) the density matrix (\ref{3.5}) is

\begin{eqnarray}\rho (\sigma)&&={\cal T}\int D\eta e^
{-2\gamma\int_{-\infty}^{\infty} dxdx'\eta(x)G^{-1}(x-x')\eta(x')}
e^{-i2\gamma\int_{\sigma_{0}}^{\sigma}dx\eta(x)[\phi(x)\otimes 1-1\otimes \phi(x)]}\nonumber\\
&&\qquad\qquad\qquad\qquad \cdot e^{-ig(2M)\int_{\sigma_{0}}^{\sigma}dx
[\phi (x):\psi^{2} (x):\otimes 1-1\otimes\phi (x):\psi^{2} (x):]}\rho ({\sigma_{0})}
\label{4.1}\end{eqnarray}

\noindent In Eq. (\ref{4.1}), ($\sigma_{0},\sigma)$ are arbitrary nonintersecting 
initial and final spacelike hypersurfaces (which will be replaced by  $(-T/2,T/2)$ for calculations), 
and $G(x-x')$ is a Lorentz scalar:

	\begin{equation}G(x-x')={1\over (2\pi)^{4}}\int dke^{ik\cdot(x-x')}\tilde G(k^{2})\label{4.2}\end{equation}
	
\noindent where $\tilde G(k^{2})$ is real and positive.  

	It is shown in Appendix \ref{app:C} that such a model is Lorentz invariant 
from both the passive point of view (i.e., the description of the evolution of the statevector 
$|\psi,\sigma_{0}>$ to $|\psi,\sigma>$ is the same from all Lorentz frames) 
and the active point of view (i.e., if $|\psi,\sigma_{0}>$ evolves into $|\psi,\sigma>$ 
then the Lorentz transform of $|\psi,\sigma_{0}>$ evolves into the Lorentz transform of 
$|\psi,\sigma>$) .  

	RCSL$_1$ is the (Markovian) model obtained with $G(x-x')=\delta (x-x')$, i.e., 
$\tilde G(k^{2})=1$. The qualitative 
collapse behavior of this relativistic model (which is the same as in all the rest of 
the models presented here) has been discussed 
extensively \cite{GGP,GP',GG',G'}, especially by Ghirardi and Grassi, and we shall 
not go over that ground in any detail here.  
The main feature is that the picture of the collapse process is 
hypersurface dependent (e.g., in the example at the end of 
section \ref{sec:CSLexample}, the spacetime region where 
the wavepacket ``there" disappears depends upon the reference frame 
from which the evolution is described).  Nonetheless, this feature is 
experimentally consistent as ``collapse time" and ``collapse location" 
are not measurable quantities, and all observers agree on what \emph{is} 
measurable, the collapse result. Some of this 
behavior was presaged in their discussion of the 
collapse postulate in relativistic quantum theory by Aharonov and Albert\cite{AA}.   
 
	We shall now discuss the calculation of the energy produced from the vacuum to (lowest) order 
$\gamma$: with no extra labor we can 
consider RCSL$_1$ together with the more general case where $\tilde G(k^{2})$ does not have to equal 1.
It is useful to introduce Feynman diagrams to discuss the perturbation 
expansion of the density matrix (\ref{4.1}).  
The diagrammatic representation of the two vertices, $2\gamma\eta(x)\phi(x)$ and 
$g(2M)\phi (x):\negthinspace\negthinspace\psi^{2} (x)\negthinspace\negthinspace:$, 
are illustrated in Figs. 1a, 1b. 

	After performing the perturbation expansion in powers of $\gamma$ and $g$, we must integrate 
over $\eta$ using, for example,  

	\begin{equation}\int D\eta e^{-2\gamma\int_{-\infty}^{\infty} dxdx'\eta(x)G^{-1}(x-x')\eta(x')}
	2\gamma \eta (z)=0\label{4.3}\end{equation}
	\begin{equation}\int D\eta e^{-2\gamma\int_{-\infty}^{\infty} dxdx'\eta(x)G^{-1}(x-x')\eta(x')}
	2\gamma \eta (y)2\gamma \eta (z)=\gamma G(y-z)\label{4.4}\end{equation}
	
\noindent   Odd powers of $\gamma \eta$ give $0$, 
while even powers give products of factors $\gamma G(x_{i}-x_{j})$, with all possible permutations of
the indices.  Thus all vertices containing $\eta$ are ``paired up," and may be regarded as 
connected by a ``noise propagator" $\gamma G(x_{i}-x_{j})$. 
The symbol for the noise propagator is a thick line, 
which is illustrated connecting two $\eta$ vertices in Fig. 1c. In momentum space, 
the noise propagator contributes a factor $\gamma \tilde G(k^{2}).$

	Because it is a density matrix we are expanding, we get terms to the left of $\rho (T_{0})$ and 
terms to the right of $\rho (T_{0})$.  In the diagrams we draw, these shall be separated by a 
vertical dotted line.  The only thing that can cross the line is a 
noise propagator.   We note that operators 
to the right of $\rho (T_{0})$ are time-reversed.  This means that the four-momenta may be 
considered as going backwards in time to the right of the dotted line  (or alternatively, we 
may consider that they are negative four-momenta going forward in time).  

	In RCSL$_1$, particles are produced from the vacuum.   
We shall now see that this may be avoided to order $\gamma$ if 
$\tilde G(k^{2})$ vanishes for timelike $k^{2}$. The diagrams 
which describe the evolution of the vacuum 
to order $\gamma$ are shown in Figs. 2a,b,c. Diagrams 2a,b
describe the decrease of the vacuum state amplitude due to $\phi$-particle creation, 
while 2c describes creation of a $\phi$-particle.  The  
contribution of these three diagrams is calculated in Appendix \ref{app:D}. 
For diagram 2c, which is connected across 
the dotted line, four-momentum goes in at the right and comes out at the left: in this way 
it is possible, from the diagrammatic point of view, to have creation of a 
$\phi$-particle out of the vacuum and still conserve four-momentum. From the physical 
point of view, of course, it is the four-momentum in the noise that generates the $\phi$-particle.  

	The result of the calculation, Eq. (\ref{D4}), is that the 
average created energy/sec-vol is proportional to  
$\tilde G(\mu^{2})$ (and infinite, because it is also proportional to $\int d{\bf k})$. Thus, 
if $\tilde G(\mu^{2})=0$, then there is no $\phi$-particle 
production out of the vacuum to order $\gamma$. Indeed, then there is no $\phi$-particle 
production out of the vacuum to order $\gamma^{n}$ (arbitrary $n$), since all $\phi$-particle 
creation diagrams not involving $g$   
consist of disconnected diagrams of type 2c, or such diagrams with the 
noise propagator replaced by alternating noise propagators and $\phi$ propagators.  

	However, there is still particle pair production out of the vacuum to order $\gamma g^{2}$.  
The relevant diagram is shown in Fig. 2d.  Its contribution is

	\begin{equation}\sim \gamma g^{2}{\tilde G[(p_{1}+p_{2})^{2}]\over [(p_{1}+p_{2})^{2}-\mu^{2}]^{2}}
	 \label{4.5}\end{equation}
	
\noindent where $p_{1}$, $p_{2}$ label the outgoing particles. To prevent this, $\tilde G(k^{2})$ 
must vanish for $k^{2}\geq (2M)^{2}$.  

	Now, we may assume that $w({\bf x}, t)$ can be coupled to any 
particle with nonzero mass \cite{PearleSquires}, e.g., an electron, 
or a neutrino (if it should turn out to have mass).  Thus the condition of 
no vacuum excitation to order $\gamma g^{2}$ leaves very little room on the 
$k^{2}\geq 0$ line where $\tilde G(k^{2})$ does not vanish. In fact, next 
consider $\phi$-particle pair production to order $\gamma g^{6}$, contributed by 
the diagram (2e).  For this to vanish, $\tilde G(k^{2})$ must vanish for 
$k^{2}\geq (2\mu)^{2}\approx(2{\rm eV})^{2}$.  Finally, suppose the particle is charged, 
so it is coupled to the electromagnetic field.  A diagram like 2e, with the 
created $\phi$-particles replaced 
by photons, shows that we must have $\tilde G(k^{2})\equiv 0$
for $k^{2}\geq 0$ to prevent photon production to order $\sim\gamma$.

	Thus to prevent particles from being created 
out of the vacuum at an infinite rate/vol to an order $\sim\gamma$ 
we must let $\tilde G(k^{2})\not = 0$ only for spacelike $k^{2}$.

\section{RCSL$_{2}$}\label{sec:RCSL2}

	Does restricting $\tilde G(k^{2})$ to be nonvanishing only for $k^{2}<0$ make collapse behavior 
untenable?  On the contrary, as we shall see, 
$\tilde G(k^{2})$ for $k^{2}\geq 0$ is irrelevant for collapse. 
 
	Consider the behavior of $G(x-x')$ in the limit $c\rightarrow \infty$:   
this is one way of looking at the nonrelativistic limit.  Use of Eq. (\ref{4.2}) 
with the $c$ dependence 
explicitly displayed results in

\begin{eqnarray}G({\bf x}-{\bf x}', t-t')&&={1\over (2\pi)^{4}}
\int d\omega d{\bf k}e^{i\omega(t-t')-i{\bf k}\cdot({\bf x}-{\bf x}')}
\tilde G({\omega ^{2}\over c^{2}}-{\bf k}^{2})\nonumber\\
&&\longrightarrow\delta (t-t') {1\over (2\pi)^{3}}
\int d{\bf k}e^{-i{\bf k}\cdot({\bf x}-{\bf x}')}\tilde G(-{\bf k}^{2}) 
\label{5.1}\end{eqnarray}

\noindent Thus it is only negative values of $k^{2}$ which are relevant 
arguments of $\tilde G$ in this limit.

	We note that the time dependence of $G$ in this limit is $\sim \delta (t-t')$, 
identical to that of 
CSL.  What about the spatial behavior of the limit (\ref{5.1})?  In the rest of this paper we 
shall make the simplest choice of spacelike spectrum 

\begin{equation}\tilde G(k^{2})=\delta (k_{0}^{2}-{\bf k}^{2}+\mu^{2})\label{5.2}\end{equation}

\noindent corresponding to a tachyon of mass 
$\mu=a^{-1}\approx 1$eV.  With this choice, Eq. (\ref{5.1}) becomes 

\begin{equation}G({\bf x}-{\bf x}',t-t')\longrightarrow\delta (t-t') {1\over (2\pi)^{2}}
{\sin [|{\bf x}-{\bf x}'|/a]\over|{\bf x}-{\bf x}'|}\label{5.3}\end{equation}

\noindent The spatial dependence of $G$ in Eq. (\ref{5.3}) 
is quite a satisfactory replacement\cite{Weber} for the Gaussian dependence (\ref{3.2b}) 
introduced by GRW and which is usually employed in CSL.  

	It is worth noting here the exact solution for $G$ of which (\ref{5.3}) 
is the $c\rightarrow \infty$
limit.  Putting (\ref{5.2}) into (\ref{4.2}) yields\cite{Feinberg} 

\begin{mathletters}\label{5.4}
\begin{eqnarray}G(x-x')&&=-{1\over 8\pi ^{2} a|x-x'|}N_{1}(|x-x'|/a)\qquad\qquad (x-x')^{2}<0\\
&&=-{1\over4\pi^{3}a|x-x'|}K_{1}(|x-x'|/a)\qquad\qquad (x-x')^{2}>0\label{5.4a, b}
\end{eqnarray}
\end{mathletters}

\noindent ($|x-x'|\equiv \sqrt{|(x-x')^{2}|}$ and $N$, $K$ are Bessel Functions). 
For spacelike $x-x'$, $G$ oscillates 
on the scale $a\approx 10^{-5}$cm (and decreases as 
$|x-x'|^{-{3\over 2}}$), while for timelike $x-x'$, $G$ decays 
exponentially with time constant $a/c\approx 10^{-15}$sec. 

	For RCSL$_1$, where $\tilde G(k^{2})=1$, the spatial scale of collapse is not set 
by G: rather it is set by the $\phi$-mass which 
governs the spatial extent of the Yukawa dressing of the particles.  
With the same mass chosen for the tachyon, this 
physical mechanism for the scale appears to be---and is---redundant: the $\phi$-particle 
is no longer needed. 
 
	To see this, compare the RCSL$_1$ diagrams in Fig. 3 with their counterpart diagrams 
in Fig. 4, belonging to what we shall call RCSL$_2$.  In each RCSL$_1$ diagram, 
two $\phi$-propagators are attached to the ends of a noise propagator, providing a factor 
$\sim \gamma \tilde G(k^{2})(k^{2}-\mu^{2})^{-2}=\gamma \delta (k^{2}+\mu^{2})(-2\mu^{2})^{-2}$.  
Thus the $\phi$-propagators just provide a numerical factor (and 
would do so even were the $\phi$ mass not $\mu$) 
which may be incorporated into $\gamma$.  Therefore the $\phi$ particles may be completely dispensed 
with. (We note that external $\phi$-lines which show up in 
higher order diagrams of RCSL$_1$, describing production, 
absorption or scattering of $\phi$ particles, are not needed for collapse behavior.) 

	The new model, RCSL$_2$, is 
described by the statevector evolution Eq. (\ref{3.1}) 
with $G(x-x')$ given by Eqs. (\ref{4.2}), (5.2) and with 
$F(x)=2M:\negthinspace\negthinspace\psi^{2}(x)\negthinspace\negthinspace:$\cite{Grassi1}.
The associated density matrix is 

\begin{equation}\rho (\sigma)={\cal T}\negthinspace\int D\eta e^
{-2\gamma\int_{-\infty}^{\infty} dxdx'\eta(x)G^{-1}(x-x')\eta(x')}
e^{-i2\gamma(2M)\int_{\sigma_{0}}^{\sigma}dx\eta(x)[:\psi^{2} (x):\otimes 1-1\otimes :\psi^{2} (x):]}
\rho ({\sigma_{0})}\label{5.5}\end{equation}

\noindent The coupling constant 
$\gamma$ is dimensionless. In the nonrelativistic limit discussed above and with neglect of 
pair creation/annihilation (replacement of 
$:\negthinspace\negthinspace\psi^{2}(x)\negthinspace\negthinspace:$ 
by $2\xi^{\dagger}(x)\xi (x)/2M$), it is 
seen that the nonrelativistic limit of RCSL$_2$ is CSL.  

\section{Density Matrix to Lowest Order in RCSL$_{2}$}\label{sec:densmatrix}
 
	The perturbation series diagrams corresponding to Eq. (\ref{5.5}) 
are built out of just particle lines,
tachyon propagators (which we formerly called noise propagators) and vertices where one 
end of a tachyon propagator meets two particle lines. 
To (lowest) order $\gamma$, the nonvanishing diagrams, which are responsible for 
collapse and which replace those in Figs. 3a,b,c, 
are shown in Figs. 4a,b,c.  We shall obtain an expression for their 
contribution here.  However,  
in order to bring out the similarity to nonrelativistic CSL, we 
shall not use diagrammatic methods.

	The density matrix (\ref{5.5}) to lowest order is 
	
\begin{eqnarray}\rho (T/2)&&=\rho (-T/2)-{1\over 2}\gamma (2M)^{2}
\int_{-{T\over 2}}^{T\over 2} dxdx'G(x-x')\nonumber\\
&&[{\cal T}:\negthinspace\negthinspace\psi^{2} (x)\negthinspace\negthinspace:
:\negthinspace\negthinspace\psi^{2} (x')\negthinspace\negthinspace:\rho (-T/2)+
\rho (-T/2){\cal T}_{\cal R}
:\negthinspace\negthinspace\psi^{2} (x)\negthinspace\negthinspace:
:\negthinspace\negthinspace\psi^{2} (x')\negthinspace\negthinspace:\nonumber\\
&&\qquad\qquad\qquad\qquad\qquad\qquad\qquad
-2:\negthinspace\negthinspace\psi^{2} (x)\negthinspace\negthinspace:
\rho (-T/2):\negthinspace\negthinspace\psi^{2} (x')\negthinspace\negthinspace:]
\label{6.1}\end{eqnarray}

\noindent (${\cal T}_{\cal R}$ is the time-reversing operator).  Utilizing 

\begin{equation}{\cal T}:\negthinspace\negthinspace\psi^{2} (x)\negthinspace\negthinspace:
:\negthinspace\negthinspace\psi^{2} (x')\negthinspace\negthinspace:
=:\negthinspace\negthinspace\psi^{2} (x')\negthinspace\negthinspace:
:\negthinspace\negthinspace\psi^{2} (x)\negthinspace\negthinspace:
+\Theta (t-t') [\psi^{2} (x),\psi^{2} (x')]\label{6.2}\end{equation}

\noindent and taking advantage of the $x\leftrightarrow x'$ symmetry 
allows writing Eq. (\ref{6.1}) as

\begin{eqnarray}\rho (T/2)=&&\rho (-T/2)-{1\over 2}\gamma (2M)^{2}
\int_{-{T\over 2}}^{T\over 2} dxdx'G(x-x')
[:\negthinspace\negthinspace\psi^{2} (x)\negthinspace\negthinspace:
[:\negthinspace\negthinspace\psi^{2} (x')\negthinspace\negthinspace:,\rho (-T/2)]]\nonumber\\
&&+{i\over 4}\gamma (2M)^{2}\int_{-{T\over 2}}^{T\over 2} dxdx'G(x-x')
\epsilon (t-t')[i[\psi^{2} (x),\psi^{2} (x')],\rho (-T/2)]\label{6.3}\end{eqnarray}

	Now we note the very important relation

\begin{equation}\int_{-{T\over 2}}^{T\over 2} dx'G(x-x')\psi_{\pm}^{2} (x')\longrightarrow 0
\qquad \hbox {as\ }T\rightarrow \infty\label{6.4}\end{equation}

\noindent where $\psi_{+}(x')$ and $\psi_{-}(x')$ are the positive and negative 
frequency parts respectively of $\psi(x')$. The expression (\ref{6.4}) vanishes because,  
in momentum space, the integrand is 
$\sim\delta (k^{2}+\mu^{2})\delta^{4}[k\pm(p_{1}+ p_{2})]$, where $p_{1}$ and $p_{2}$ are timelike, 
and the sum of two timelike four-momenta cannot equal a spacelike 
four-momentum. Eq. (\ref{6.4}) guarantees that no 
pair production out of the vacuum arises from the first perturbation term in 
Eq. (\ref{6.3}), as it permits $:\negthinspace\negthinspace\psi^{2}(x)\negthinspace\negthinspace:$ 
to be replaced by $2\psi_{-}(x)\psi_{+}(x)$. 
Eq. (\ref{6.4}) will turn out to be crucial in obtaining 
no particle creation out of the 
vacuum to any order in the RCSL model discussed in the next section, and crucial 
for eliminating the need for renormalization.    

	There is no pair production out of the vacuum arising from the 
second perturbation term in Eq. (\ref{6.3}) either, since its pair 
production/annihilation part has the form

\[\int_{-{T\over 2}}^{T\over 2} 
dxdx'\psi_{\pm} (x)\psi_{\pm} (x')F(x-x')\longrightarrow 0
\qquad \hbox{as\ }T\rightarrow \infty\]
 
\noindent (because in momentum space the integrand 
is $\sim \delta^{4}(p+p')=0$).  Thus we finally obtain, for large $T$, 

\begin{eqnarray}&&\rho (T/2)=\rho (-T/2)-2\gamma (2M)^{2}
\negthinspace\negthinspace\negthinspace
\int_{-{T\over 2}}^{T\over 2} dxdx'G(x-x')
[\psi_{-}(x)\psi_{+}(x),[\psi_{-}(x')\psi_{+}(x'),\rho (-T/2)]]\nonumber\\
&&+i2\gamma (2M)^{2}\negthinspace\negthinspace\negthinspace
\int_{-{T\over 2}}^{T\over 2} dxdx'G(x-x')
\epsilon (t-t')<0|i[\psi (x),\psi (x')]|0>[\psi_{-} (x)\psi_{+}(x'),\rho (-T/2)]
\label{6.5}\end{eqnarray}
           
	The first perturbation term in Eq. (\ref{6.5}) describes collapse. (It corresponds to 
diagrams 4b,c as well as the real part 
of diagrams 4a.) Some of its consequences will be 
obtained in section \ref{sec:Results}. 
It is quite similar in form to the comparable nonrelativistic CSL expression 
(obtained from the perturbation expansion of Eq. (\ref{2.7})):

\begin{eqnarray}\rho (T/2)=\rho (-T/2)&&-{\lambda\over 2}
\int_{-{T\over 2}}^{T\over 2} dxdx'\delta (t-t')
e^{-{1\over 4a^{2}}({\bf x}-{\bf x}')^{2}}\nonumber\\ 
&&\qquad\cdot[\xi^{\dagger}({\bf x},t)\xi({\bf x},t),
[\xi^{\dagger}({\bf x}',t)\xi({\bf x}',t),\rho (-T/2)]] \label{6.6}\end{eqnarray}

	 The second perturbation term in Eq. (\ref{6.5}) is a unitary 
(noncollapse) evolution. This term comes from the imaginary part of the 
diagrams 4a. It is the divergent, imaginary, part of the self-energy diagrams 4a.  
While it may be removed by mass renormalization, we shall remove it a different way.

\section{RCSL}\label{sec:RCSL}

	The RCSL$_{2}$ choice of tachyonic $\tilde G (k^{2})$ means 
there is no particle production out of the vacuum in order $\gamma$. However, 
there is pair production 
out of the vacuum in order $\gamma^{2}$ via the diagram in Fig. 5. The culprit 
is the off-shell four-momentum (i.e., $p^{\nu}$ can take on any value) of the Feynman 
propagator.  If the propagator were on-shell (i.e., $p^{2}=M^{2}$), energy conservation at the 
lower vertices in Fig. 5 would make the contribution of the diagram vanish for the 
same reason that (\ref{6.4}) vanishes.  

	We shall now see how to achieve this 
so as to eliminate vacuum excitation to 
this order, and indeed to all orders.  This 
can be accomplished if we focus on an apparently unrelated task, eliminating 
the second perturbation 
term in Eq. (\ref{6.5}).  It is not unrelated however, because this term 
depends upon $\epsilon (t-t')<0|i[\psi (x),\psi (x')]|0>$ whose Fourier transform is 
${\cal P}(p^{2}-m^{2})^{-1}$, the off-shell (imaginary) part of the Feynman propagator.   

	Consider the statevector and density matrix 
evolution equations (\ref{3.4}), (\ref{3.5}) where we replace 
$F(x)=2M:\negthinspace\negthinspace\psi^{2}(x)\negthinspace\negthinspace:$ by  
$F(x)=2M:\negthinspace\negthinspace\psi^{2}(x)\negthinspace\negthinspace:+ F_{1}(x)$ with 
	
\begin{equation}F_{1}(x)\equiv 
	\gamma(2M)^{2}\int_{T_{0}}^{t}dx' \eta (x')i[\psi^{2}(x),\psi^{2}(x')]\label{7.1}\end{equation}
	
\noindent (it is to be understood that the time--ordering operation is to 
treat the operator $F_{1}(x)$ as a 
function of $t$).  When this new expression for $F(x)$ is put into the density matrix 
(\ref{3.5}), $F_{1}(x)$ adds to the lowest order expression (\ref{6.5}) a contribution 
which is the negative of the last term in (\ref{6.5}).  

	What has happened is this: where 
${\cal T}:\negthinspace\negthinspace\psi^{2}(x)\negthinspace\negthinspace:
:\negthinspace\negthinspace\psi^{2}(x')\negthinspace\negthinspace:$ appears in the lowest order 
Feynman diagrams, $F_{1}(x)$ adds $-\Theta (t-t')[\psi^{2}(x),\psi^{2}(x')]$ to it.  Thus, as 
can be seen from Eq. (\ref{6.2}), {\it the extra 
term $F_{1}(x)$ effectively removes the time--ordering operation}.  

	It also turns out (we shall spare the reader the details) that use of (\ref{7.1}) also removes 
the $\cal T$ operator in the next order, 
from the expression for the pair production diagram of Fig. 5. 
But, because of Eq. (\ref{6.4}),

\begin{equation}\int dxG(x-x'):\negthinspace\negthinspace\psi^{2}(x)\negthinspace\negthinspace:=
\int dxG(x-x')2\psi_{-}(x)\psi_{+}(x)\label{7.2}\end{equation}

\noindent replacing 
${\cal T}:\negthinspace\negthinspace\psi^{2}(x)\negthinspace\negthinspace:
:\negthinspace\negthinspace\psi^{2}(x')\negthinspace\negthinspace:$ by
$:\negthinspace\negthinspace\psi^{2}(x)\negthinspace\negthinspace:
:\negthinspace\negthinspace\psi^{2}(x')\negthinspace\negthinspace:$ 
has the effect, for the diagram of Fig. (5), of replacing 
the term ${\cal T}:\negthinspace\negthinspace\psi^{2}(x)\negthinspace\negthinspace:
:\negthinspace\negthinspace\psi^{2}(x')\negthinspace\negthinspace:$, which can create particle 
pairs out of the vacuum, by the term 
$4\psi_{-}(x)\psi_{+}(x)\psi_{-}(x')\psi_{+}(x')$ which 
annihilates the vacuum.  Adding $F_{1}(x)$ has made the contribution of 
the diagram of Fig. 5 vanish: there is now no vacuum excitation to 
order $\gamma^{2}$.

	It is now clear that a way to remove vacuum excitation to all orders is to  
keep the RCSL$_{2}$ statevector evolution except {\it remove the time ordering operator} ${\cal T}$:

\begin{equation}|\psi ,\sigma>=\int D\eta e^
{-\gamma\int_{-\infty}^{\infty} dxdx'\eta(x)G^{-1}(x-x')\eta(x')}
e^{i\int_{\sigma_{0}}^{\sigma}dx\eta(x)[w(x)-2\gamma (2M):\psi^{2}(x):]}
|\psi ,{\sigma_{0}}>\label {7.3}\end{equation}

\noindent  This model I shall call RCSL. 

	When the time ordering operator ${\cal T}$ is removed, the form (\ref{3.1}) 
and its Fourier transform form (\ref{3.4}) are 
generally no longer equal.  The Fourier transform form, (\ref{7.3}), has 
to be chosen to define RCSL because only 
then is the integrated probability density (\ref{2.2}) equal to 1. The RCSL 
density matrix is (\ref{5.5}) 
without the $\cal T$--operation: 

\begin{equation}\rho (\sigma)=\int D\eta e^
{-2\gamma\int_{-\infty}^{\infty} dxdx'\eta(x)G^{-1}(x-x')\eta(x')}
e^{-i2\gamma(2M)\int_{\sigma_{0}}^{\sigma}dx
\eta(x)[:\psi^{2}(x):\otimes 1
-1\otimes :\psi^{2}(x):]}
\rho ({\sigma_{0})}\label{7.4}\end{equation}

\noindent It is easily seen that the trace of (\ref{7.4}) (the integrated probability density) 
is 1.  

	When (\ref{7.4}) is expanded in a perturbation series and (\ref{6.4}) is employed, 
the result for large $T$ is 

\begin{eqnarray}\rho (T/2)&&=\sum_{n=0}^{\infty} {(-16\gamma M^{2})^{n}\over (2n)!}
\int_{-T/2}^{T/2} dx_{1}...dx_{2n}
\sum_{C}G(x_{i_{1}}-x_{i_{2}})...G(x_{i_{2n-1}}-x_{i_{2n}})\nonumber\\
&&\qquad\qquad\cdot 
[\psi_{-}(x_{1})\psi_{+}(x_{1}),[...[\psi_{-}(x_{2n})
\psi_{+}(x_{2n}),\rho(-T/2)]...]]\label{7.5}\end{eqnarray}

\noindent (the summation is over the $(2n-1)!!$ pairing combinations of the integers 1,...2n). Thus, 
two features, choice of tachyonic $\tilde G(k^{2})$ and removal of the time-ordering operation, 
has resulted in a model which is  
free from vacuum excitation, dependent as it is on 
operators only in the combination  $\psi_{-}(x)\psi_{+}(x)$.

	Indeed, it is completely finite: there is no need for renormalization.   

	RCSL is Lorentz invariant:  the arguments in 
Appendix \ref{app:C} may just as well have been applied there to the 
Fourier transform representation of the 
statevector (the only additional information needed is that $\eta(x)$ transforms as a 
Lorentz scalar).  With or without the ${\cal T}$-- 
operation, operators and variables transform as stated, so the demonstration 
goes through as before. 

	Incidentally, it 
may be remarked that one would not have a 
Lorentz invariant statevector or density matrix if such products as 
${\cal T}\psi_{-}(x)\psi_{+}(x')$ appears in them (the time-ordering is not Lorentz invariant since
$\psi_{-}(x)$ and $\psi_{+}(x')$ do not commute for spacelike $x-x'$) but 
$\psi_{-}(x)\psi_{+}(x')$ does appear in RCSL 
expressions  and it is Lorentz invariant.  Indeed, one could 
replace $:\negthinspace\negthinspace\psi^{2}(x)\negthinspace\negthinspace:$ in 
Eqs. (\ref{7.3}), (\ref{7.4})  by $2\psi_{-}(x)\psi_{+}(x)$) without affecting the S-matrix behavior, 
but also without violating Lorentz invariance. So far I have no reason for choosing 
one form over the other.

	It is also worth remarking that removing the time-ordering operation in a normal field 
theory with an interaction of the form $\psi^{2}(x)\phi(x)$ is 
trivial (if the mass of the $\phi$ particle is less than twice the mass of the $\psi$) since, if 
all the particles are on-shell, 
energy-momentum conservation at each vertex forbids any evolution.  It is only 
with the $\phi$ particle replaced by a tachyon that evolution is possible.   

	RCSL describes collapse for an individual statevector.  For the usual test, we set $H=0$ 
in Eq. (\ref{7.3}) so that the unitary evolution does not interfere with the collapse evolution.  
When we replace $\psi_{-}(x)\psi_{+}(x)$ by $\psi_{-}(\bf{x}, 0)\psi_{+}(\bf{x}, 0)$, then all 
operators commute. It is then irrelevant whether or not a time-ordering operator is 
present and the demonstration of individual 
collapse is that in Appendix \ref{app:A}.
	
	One may represent the terms in the expansion (\ref{7.5}) by Feynman diagrams.  
The usual off-shell Feynman propagator for the 
particle, $<0|{\cal T}\psi(x)\psi(x')|0>$, is replaced by the on-shell positive-energy 
propagator $<0|\psi(x)\psi(x')|0>$.  Since the tachyon propagator is on-shell as well, 
we may think of all diagrams as representing real 
(as opposed to virtual) particles emitting or absorbing real tachyons. 
It is easy to calculate from conservation of energy-momentum
that a particle of mass M at rest emits a 
negative energy tachyon of total energy $-\mu^{2}/2M$, and goes off with kinetic energy $\mu^{2}/2M$.  
In a sense, \emph{the GRW spontaneous ``hits" in CSL may 
be thought of as replaced by the spontaneous emission/absorption 
of tachyons in RCSL as the cause of collapse}.

\section{Locality and Nonlocality in RCSL}\label{sec:Locality}

	We have removed the time-ordering operation as one way to obtain a finite theory.  
One expects that it will lead to nonlocal behavior. 
In this section we explore what that behavior is and how bad it is.
 
\subsection{No Time-Ordering}\label{sec:Notime}

		In Appendix \ref{app:E1} it is shown how to write 
$\exp -i\int dtV(t)$ in the form ${\cal T}\{\exp -i\int dtH_{\mbox{eff}}(t)\}$.  As 
pointed out there, $V$ and $H_{\mbox{eff}}$ are identical to order $\gamma$, but to 
order $\gamma^{2}$ and higher, $H_{\mbox{eff}}$ has a nonlocal part which 
transports a particle from $x$ to $x'$.  $x$ and $x'$
are timelike related, but the transport can be in 
the backwards as well as the forwards time direction.  
Higher order terms describe a sequence of backwards and forwards motions in time between 
$x$ and $x'$. 
  
	Another way to see that backwards-in-time motion is characteristic of no time-ordering 
is as follows.  The standard evolution of a statevector in the interaction picture is      

\begin{eqnarray}{\cal T}&&\{ e^{-i\int_{0}^{T}dtV(t)}\} |L_{0},R_{0}>= 
\sum_{n}(-i)^{n}e^{iH_{0}T}\int_{0}^{T}dt_{n}\cdot\cdot\cdot
\int_{0}^{t_{3}}dt_{2}\int_{0}^{t_{2}}dt_{1}\cdot\nonumber\\
&&\qquad\qquad
e^{-iH_{0}(T-t_{n})}V(0)\cdot\cdot\cdot 
e^{-iH_{0}(t_{3}-t_{2})}V(0)
e^{-iH_{0}(t_{2}-t_{1})}V(0)e^{-iH_{0}t_{1}}|L_{0},R_{0}>\label{8.1}\end{eqnarray}

\noindent In (\ref{8.1}), particles evolve freely 
for $t_{1}$ sec, change momenta, evolve freely for $(t_{2}-t_{1})$ sec, 
change momenta, etc., until time $T$. All the motion is forward in time. (The $\exp{iH_{0}T}$ 
at the left end of the integrals is an artifact of the interaction picture, not a true 
backwards-in-time motion, and is removed in the Schrodinger picture.) 
Now consider this evolution \emph{without} the time-ordering operation: 
	
\begin{eqnarray}&&\ e^{-i\int_{0}^{T}dtV(t)}|L_{0},R_{0}>=
\sum_{n}\frac{(-i)^{n}}{n!}e^{iH_{0}T}\int_{0}^{T}dt_{n}
e^{-iH_{0}(T-t_{n})}V(0)e^{-iH_{0}t_{n}}\cdot\cdot\cdot\nonumber\\
&&\qquad\qquad e^{iH_{0}T}\int_{0}^{T}dt_{2}
e^{-iH_{0}(T-t_{2})}V(0)e^{-iH_{0}t_{2}}e^{iH_{0}T}\int_{0}^{T}dt_{1}
e^{-iH_{0}(T-t_{1})}V(0)e^{-iH_{0}t_{1}}|L_{0},R_{0}>\label{8.2}\end{eqnarray}
  
\noindent In (\ref{8.2}), particles evolve freely for $t_{1}$ sec, change momenta, 
evolve freely for $(T-t_{1})$ sec, and then evolve freely \emph{backwards} in time 
for $T$ sec, and then do it all over again repeatedly. 

	Yet a third way to see this 
is to write (\ref{8.2}) in a different 
way, e.g., the second order term may be written as 
 
\begin{eqnarray*}&&\int_{0}^{T}dt_{2}V(t_{2})\int_{0}^{T}dt_{1}V(t_{1})= 
e^{iH_{0}T}\Biggl[\int_{0}^{T}dt_{2}e^{-iH_{0}(T-t_{2})}V(0)\int_{0}^{t_{2}}dt_{1}
e^{-iH_{0}(t_{2}-t_{1})}V(0)e^{-iH_{0}t_{1}}\\
&&\qquad\qquad+\int_{0}^{T}dt_{2}e^{-iH_{0}(T-t_{2})}V(0) \int_{t_{2}}^{T}dt_{1}
e^{iH_{0}(t_{2}-t_{1})}V(0)e^{-iH_{0}t_{1}}\Biggr]\end{eqnarray*}

\noindent as the sum of a time-ordered evolution, and of an evolution which goes freely backwards in 
time between the two interactions.  

	A sequence of backward and forward motions in time can lead to spacelike (superluminal) 
transport, but it doesn't {\it have} to.  
 
\subsection{Particle Motion In RCSL}\label{sec:ParticleMotion}

	In this subsection we shall see, for particles of mass $M>>\mu$, 
spacelike transport only occurs at a very high order of perturbation theory, and so has 
a negligibly small probability.
 	 
	 The motion of a particle can be envisioned as a sequence of 
emissions/absorptions of tachyons.  In Appendix \ref{app:E2}, 
the evolution over time $T$ of the wavefunction of a  
particle initially localized in position and momentum is calculated to 
lowest order (diagram 4c).  
For example, if the particle is initially at rest (i.e., 
described by a wavepacket centered around momentum 0 and, 
say, around the position ${\bf x}\approx 0$), 
its wavepacket spreads a distance $(\mu/M)cT$ in the 
direction of the momentum ${\bf p}=\mu c{\hat{\bf p}}$ which it gains from an emitted tachyon. 
In addition to this spread, the resulting wavefunction 
acquires the momentum ${\bf p}$ (acquires a phase factor $\exp i{\bf p}\cdot{\bf x}$). 

	This is precisely the same as  
the behavior of a collection of classical particles,  
each of which starts at ${\bf x}=0$ at $t=0$ but,
at a random time $t$ in the interval $(0,T)$, receives an impulse resulting in 
momentum ${\bf p}$: a particle receiving the impulse at $t=0$ travels
the farthest distance, $(\mu/M)cT$.

	The backwards 
in time evolution $\exp iH_{0}T$ then carries the wavefunction a distance $(\mu/M)cT$
in the opposite direction from ${\hat{\bf p}}$ so that an elongated wavefunction stretching from 
$\approx -(\mu/M)cT{\hat{\bf p}}$ to the origin and describing a particle with momentum ${\bf p}$, 
awaits the next evolution. From the point of view of the classical particle picture, 
the particle which received its impulse at $t=0$ ends up where it started, at ${\bf x}=0$, 
while the particle which received its impulse at $t=T$ ends up at $-(\mu/M)cT{\hat{\bf p}}$.  
	
	The tachyons are, however, emitted in random directions so the complete 
wavefunction arising from diagram 4c is spread over a sphere of radius $(\mu/M)cT$.  
Thus, at each vertex the wavefunction spread is less than the width $cT$of the lightcone. 
One might think that the lightcone 
width would be exceeded if the particle initally moves with speed 
near $c$ so it is initially close to 
the lightcone, but that is not so: at relativistic speeds the spread is Lorentz contracted, and 
the wavefunction remains within the lightcone.

Now consider a higher order diagram, e.g., the iteration of fig 4c, a so-called ``ladder diagram" 
(other diagrams of the same order with more complicated configurations engender no 
more spread than this one).  At each vertex there is a further spread. One might think that 
the situation of largest spread occurs if each subsequent evolution 
described by the succeeding integrals in (\ref{8.2}) involves 
a tachyon emitted/absorbed in the same direction so that the particle is 
accelerated in the same direction in each order. However, as shown in Appendix \ref{app:E2}, 
each subsequent spread is 
slightly less than the previous one (precisely the same as the Lorentz transformed shrinking of 
the spread due to increased velocity of the particle)   
and the net spread after 
an arbitrary number of such interactions is always within the light cone. 

	However, there \emph{are} scenarios for a sequence of tachyon interactions which result in 
a spread outside the lightcone. Here is the one that spreads the 
most. It can be understood from the classical particle picture.  
Consider two successive evolutions of a particle initially at rest. 
In the first evolution, the particle starts at rest at the origin, and 
gets its momentum ${\bf p}$ at the \emph{latest} possible time, $t=T$, so that after 
travelling backward-in-time for time $T$ it is located at $ -(\mu/M)cT{\hat p}$.  In the second 
evolution, the particle gets the opposite momentum ${-\bf p}$ at the \emph{earliest} 
possible time, $t=0$.  
Thus, the particle comes to rest at location $ -(\mu/M)cT{\hat p}$, and stays there for the 
rest of that evolution.  The net effect of these two evolutions is 
that the particle ends up at rest at $ -(\mu/M)cT{\hat p}$, and if each successive pair of 
evolutions is the same, after more than $2M/\mu$ evolutions the particle wavefunction 
is outside the lightcone. But, this is a very improbable.  The first reason is that the  
phase space volume in which the succession of tachyons put their momentum in 
the right directions is small.  The second reason is that  
it occurs at a very high order of 
perturbation theory ($\approx 10^{6}$ for an electron, $(\approx 10^{9}$ for a nucleon). 
The wavefunction arising from the set of ladder diagrams is essentially that of 
a random walk of spread $\approx (\mu/M)cT$ (faster than the usual random walk $\sim\sqrt{T}$), 
wuth a very small tail outside the lightcone $cT$.  

	To summarize, in the time-ordered situation, particle wavefunctions rigorously stay within 
their lightcones: in the no time-ordered situation this is not so, but it is true 
to a very high probability.  This is what we need in the next subsection to show that the model 
is local to a very high probability. 

\subsection{Locality and RCSL}\label{sec:LocalityRCSL} 
	
	First let's review the locality criteria as manifested in a time-ordered evolution.  
	
Consider separated spatial regions $R_{0}$ and $L_{0}$ and a (possibly 
entangled) initial state $|L_{0},R_{0}>$ 
with some particles in $R_{0}$ and the rest in $L_{0}$.  
Call $R$ and $L$ the spacetime four-volume bounded by the
lightcones emanating respectively from $R_{0}$ and $L_{0}$ over the time interval $T$.   
Call $R_{t}$ and $L_{t}$ the 
intersections of $R$ and $L$ with the spacelike hyperplane $t$ for $0\leq t\leq T$.  
Suppose that $R_{T}$ and $L_{T}$ 
do not overlap. The standard evolution of a statevector in the interaction picture is 

\begin{eqnarray}{\cal T}&&\{ e^{-i\int_{0}^{T}dtV(t)}\} |L_{0},R_{0}>=\nonumber\\
&&\qquad\sum_{n}(-i)^{n}e^{iH_{0}T}\int_{0}^{T}dt_{n}\cdot\cdot\cdot
\int_{0}^{t_{3}}dt_{2}\int_{0}^{t_{2}}dt_{1}\nonumber\\
&&\qquad\qquad
e^{-iH_{0}(T-t_{n})}V(0)\cdot\cdot\cdot 
e^{-iH_{0}(t_{3}-t_{2})}V(0)
e^{-iH_{0}(t_{2}-t_{1})}V(0)e^{-iH_{0}t_{1}}|L_{0},R_{0}>\label {8.3}\end{eqnarray}

\noindent It often occurs that $V(t)|L,R>=[V_{L}(t)+V_{R}(t)]|L,R>$,  
where $|L,R>$ is any state with particles only residing within $R$ and $L$ and 
$V_{L}(t)$, $V_{R}(t)$ act only on particles within their respective lightcones. 
This requires that the free evolution $\exp -iH_{0}t$ moves particles within their 
lightcones.  This is automatic in a local relativistic quantum field theory.  
In nonrelativistic quantum mechanics this 
needs an appropriately chosen initial wavefunction (e.g., wavepacket width $>>M^{-1}$). 
It also requires $V(0)$ to give no interaction between particles in $R$ and particles in $L$.    
This is automatic in a local relativistic quantum field theory. 
In nonrelativistic quantum mechanics this is  
satisfied for a potential that is shorter range than the minimum 
separation between $R_{T}$ and $L_{T}$. 
  
	If this is the case, then taking the trace over particle states in $R_{T}$ 
results in 
\begin{mathletters}\label{8.4}
\begin{eqnarray} \mbox{Tr}_{R}
\{\rho (T)\} &&=\mbox{Tr}_{R}\{{\cal T}[ e^{-i\int_{0}^{T}dtV_{R}(t)}]
{\cal T}[ e^{-i\int_{0}^{T}dtV_{L}(t)}]|L_{0},R_{0}>\nonumber\\
&&\qquad \cdot<L_{0},R_{0}|{\cal T_{R}}[e^{i\int_{0}^{T}dtV_{L}(t)}]
{\cal T_{R}}[ e^{i\int_{0}^{T}dtV_{R}(t)}]\}\label{8.4a}\\
&&={\cal T}[e^{-i\int_{0}^{T}dtV_{L}(t)}]
\mbox{Tr}_{R}\{|L_{0},R_{0}><L_{0},R_{0}|\} 
{\cal T_{R}}[ e^{i\int_{0}^{T}dtV_{L}(t)}]\label{8.4b}\end{eqnarray}
\end{mathletters}

\noindent This is the hallmark of locality: that no evolution in 
$R$  has any effect on what happens in $L$. In particular, in an 
interacting theory, this means that no manipulation by an apparatus in $R$ 
during the time interval $(0,T)$ has any effect on $L$.  In a noninteracting theory 
like the present one this interpretation is not useful (apparatuses require interaction) 
but the requirement (\ref{8.4b}) stands.  For example, if the initial statevector is
the direct product $|L_{0}>|R_{0}>$, it means that the evolution in $L$ is
independent of whatever state is initially in $R$.

	Now consider the non time-ordered evolution.  
In the the previous subsection we have shown that a particle initially at 
the origin, although it may spread in each successive perturbation order, stays within 
the cylindrical four-volume of spatial radius $cT$ and time interval $0\leq t\leq T$ 
with a probability that is very close to 1. Suppose this were true with probability 1.  
Then the effect of  
$N(x)=\psi_{-}(x)\psi_{+}(x)$ in the perturbation series of (\ref{7.5}) could be written as 

\begin{eqnarray}\rho (T)&&=\sum_{n=0}^{\infty} {(-16\gamma M^{2})^{n}\over (2n)!}
\int_{T_{0}}^{T} dx_{1}...dx_{2n}
\sum_{C}G(x_{i_{1}}-x_{i_{2}})...G(x_{i_{2n-1}}-x_{i_{2n}})\nonumber\\
&&\qquad\qquad\cdot 
[N_{L}(x_{1})+N_{R}(x_{1}),[...[N_{L}(x_{2n})+N_{R}(x_{2n}),|L_{0},R_{0}>
<L_{0},R_{0}|]...]]\label{8.5}\end{eqnarray}

\noindent where $N_{L}({\bf x},t)$ has 
$({\bf x},t)$ lying in the cylindrical four-volume 
surrounding the $L$-particle (forwards and backwards-in-time) world-lines, 
i.e., a region of spatial extension 
$L_{T}$ at each time $0\leq t\leq T$, and similarly for $N_{R}({\bf x},t)$.  Call these 
cylindrical volumes $L'$ and $R'$. $N_{L}({\bf x},t)$ would only act on 
particles in  $L'$ and the result of the action would leave the particle in $L'$, 
and similarly for $N_{R}({\bf x},t)$. Thus, although 
$N_{L}(x)$ and $N_{R}(x')$ do not commute for all $x,x'$ nonetheless, in   
the integrand of (\ref{8.5}), one may utilize 

\begin{eqnarray*}&&[N_{L}(x_{i})+N_{R}(x_{i}),[N_{L}(x_{j})+N_{R}(x_{j}),..]]=
[N_{L}(x_{i}),[N_{L}(x_{j}),..]]+[N_{R}(x_{i}),[N_{L}(x_{j}),..]]\nonumber\\
&&\qquad\qquad\qquad\qquad\qquad\qquad\qquad 
+[N_{R}(x_{j}),[N_{L}(x_{i}),..]]+[N_{R}(x_{i}),[N_{R}(x_{j}),..]]\end{eqnarray*}

\noindent to put all $N_{R}$ commutators to the outside of the $N_{L}$ commutators.  
After relabelling of indices and summing of the series one obtains 

\begin{eqnarray}&&\rho (T)=\int D\eta e^
{-2\gamma\int_{-\infty}^{\infty} dxdx'\eta(x)G^{-1}(x-x')\eta(x')}
e^{-i2\gamma(2M)\int_{T_{0}}^{T}dx
\eta(x)[N_{R}(x)\otimes 1
-1\otimes N_{R}(x)]}\nonumber\\ 
&&\qquad\qquad\qquad\qquad \cdot e^{-i2\gamma(2M)\int_{T_{0}}^{T}dx
\eta(x)[N_{L}(x)\otimes 1
-1\otimes N_{L}(x)]}
|L_{0},R_{0}>
<L_{0},R_{0}|\label{8.6}\end{eqnarray}

	It immediately follows from (\ref{8.6}) that 
	
\begin{eqnarray}\mbox{Tr}_{R}\{\rho (T)\}&&=\int D\eta e^
{-2\gamma\int_{-\infty}^{\infty} dxdx'\eta(x)G^{-1}(x-x')\eta(x')}\nonumber\\
&&\cdot e^{-i2\gamma(2M)\int_{T_{0}}^{T}dx
\eta(x)[N_{L}(x)\otimes 1
-1\otimes N_{L}(x)]} \mbox{Tr}_{R}\{|L_{0},R_{0}>
<L_{0},R_{0}|\}\label{8.7}\end{eqnarray}

\noindent which, as in Eq. (\ref{8.4}), is the hallmark of locality.

	While Eq. (\ref{8.7}) is not strictly true in RCSL, the terms by which it deviates are 
of very small magnitude, so we say that RCSL is local to a high degree of accuracy.  

	It is worth remarking 
that if all particles are distinguishable (so there are as many quantum fields as particles, 
and $N(x)$ is replaced by the sum $\sum_{k} N_{k}(x)$ in all expressions) then (\ref{8.7}) 
is exact if one traces over the identifiable particles which start in $R$, 
no matter where they evolve to. At the ensemble level, 
this is basically a theory of 
particles which do not interact among themselves (the $N_{k}(x)$ all commute) in spite of 
the tachyonic interaction at the individual level.   
   
	It is perhaps also worth remarking that, with a local theory, there 
can be no causal loops (e.g., no sending messages to kill one's father before one is born), 
even though there are tachyonic interactions.  The essential reason is that the tachyons 
are emitted spontaneously: one cannot send a message when one does not have control 
of the messenger.  This is the same reason why one cannot communicate superluminally by 
making one part of an entangled separated system collapse: one 
cannot force a particular outcome of 
the collapse because one does not have control of the tachyonic noise field.

\section{Some Relativistic Results}\label{sec:Results}

	Now that we have a finite relativistic collapse model, we may use it.  
To illustrate, we shall present two calculations 
to order $\gamma$, using (\ref{7.5}):

\begin{equation}\rho (T/2)=\rho (-T/2)-2\gamma (2M)^{2}\negthinspace\negthinspace\negthinspace
\int_{-{T\over 2}}^{T\over 2} dxdx'G(x-x')
[\psi_{-}(x)\psi_{+}(x),[\psi_{-}(x')\psi_{+}(x'),\rho (-T/2)]]\label{9.1}\end{equation}

	First we consider the energy increase of $n$ free particles 
due to collapse.  In Appendix \ref{app:F} we calculate 
$\bar H(T/2)\equiv \hbox {Tr}H\rho (T/2)$, with the result  

\begin{equation}\bar H(T/2)=\bar H(-T/2)+{1\over 2\pi^{2}}n\gamma T {\mu^{3}\over M}
\sqrt{1+({\mu\over 2M})^{2}}\label{9.2}\end{equation}

\noindent This may be compared with the CSL energy increase 
which follows from Eq. (\ref{2.7}):
 
\begin{equation}\bar H(T/2)=\bar H(-T/2)+{3\over 4}n\lambda T {\mu^{2}\over M}\label{9.3}\end{equation}

\noindent (using $a^{-1}=\mu$). They are identical 
(for $M>>\mu$, apart from numerical factors) if we identify 
$\gamma=\lambda \mu^{-1}$: $\gamma \approx 10^{-32}$, with GRW's choice of parameter values.  
How could this small a dimensionless number be 
accounted for?  We note, following Diosi\cite{Diosi} that $GM_{N}^{2}\approx 10^{-38}$ with
$M_{N}=$ proton or neutron mass. We may set 

\begin{equation}\gamma=\kappa GM_{N}^{2}\label{9.4}\end{equation}

\noindent with $\kappa\approx 10^{6}$ for the GRW choice, while Diosi has suggested $\kappa\approx 1$.

	The nonrelativistic energy increase (\ref{9.3}) may be thought of as arising from the 
GRW ``hitting" process. This narrowing of a particle's 
wavefunction to distance $\approx \mu^{-1}$ on the 
time scale $\lambda^{-1}$ imparts to the particle 
a momentum $\sim \mu$ and thus an energy $\sim \mu^{2}/2M$. 
Likewise, the relativistic energy 
increase (\ref{9.2}) may be thought of as the consequence of 
a free particle suddenly emitting a tachyon, as represented in the diagram 
(4c). A particle at rest, which emits a negative energy 
tachyon, gains energy $\mu^{2}/2M$. 

	According to Eq. (\ref{9.2}), if the collapse 
works on particles of any mass, the mean energy imparted to zero mass 
particles becomes infinite $\sim M^{-2}$. However, experiments\cite{PearleSquires,Collett,PRCA}  
indicate that particles should be coupled to $w({\bf x},t)$ via their mass.  To take account of 
different types of particles, in (\ref{7.3}), (\ref{7.4}) one should effect the replacement 

\begin{equation}\gamma (2M):\negthinspace\negthinspace\psi^{2}(x)\negthinspace\negthinspace:\longrightarrow 
\gamma \sum_{i}{M_{i}\over M_{N}}(2M_{i}):
\negthinspace\negthinspace\psi_{i}^{2}(x)\negthinspace\negthinspace:\label{9.5}\end{equation}

\noindent ($\psi_{i}(x)$ is the field of the particle with mass $M_{i}$).  Then 
Eq. (\ref{9.2}) becomes, for the ith particle type, 

\begin{equation}\bar H(T/2)-\bar H(-T/2)={1\over 2\pi^{2}}n\gamma T {\mu^{3}M_{i}\over M_{N}^{2}}
\sqrt{1+({\mu\over 2M_{i}})^{2}}
\longrightarrow{1\over (2\pi )^{2}}n\gamma T {\mu^{4}\over M_{N}^{2}}\label{9.6}\end{equation}

\noindent in the limit $M_{i}\longrightarrow 0$. This zero-mass 
rate is $\approx (\mu/M_{N})\approx 10^{-9}$ 
times as big as the rate (\ref{9.2}) for a proton. Thus, even 
if (\ref{9.6}) were to apply to photons or 
(zero mass) neutrinos, of which the universe contains $\approx 10^{8}$ times as many as 
protons, the energy increase in the universe due to collapse of these particles would be
comparable to the energy increase in the universe due to the protons alone. However, in the strict 
zero mass limit, with use of (\ref{9.5}), the probability/sec of the tachyon reaction 
vanishes.   Such a never--occurring process, even one which produces an infinite 
energy (so the average is (\ref{9.6})) has little physical significance.  

	Now we turn to consideration of the collapse rate for a single particle in the state 
$\alpha|L>+\beta|R>$, where $|L>$ and $|R>$ are reasonably well--localized states separated 
by a great distance $>>cT$, but otherwise identical.  The off--diagonal matrix element 
$<L|\rho(T/2)|R>$ is calculated in Appendix \ref{app:G}.  The result \ref{G6}, applied to the ith 
type of particle, is

\begin{equation}<L|\rho(T/2)|R>=\alpha\beta^*\Bigl[ 1-{2\over \pi^{2}}\gamma T{\mu M_{i}^{3}\over M_{N}^{2}}
\sqrt{1+({\mu\over 2M_{i}})^{2}}\int {d{\bf p}\over 2E}|\Psi({\bf p})|^{2}\Bigr]\label{9.7}\end{equation}
  
 \noindent where $\Psi({\bf p})$ is the particle's normalized wavefunction 
 in momentum space. ((\ref{9.7}) 
arises from the ``self--energy" diagrams 4a.)  The comparable CSL result, 
obtainable from (\ref{2.7}), is: 
 
\begin{equation}<L|\rho(T/2)|R>=\alpha\beta^*
\Bigl[ 1-\lambda T{M_{i}^{2}\over M_{N}^{2}}\Bigr]\label{9.8}\end{equation}
 
 \noindent They are identical 
 (for $M_{i}>>\mu$, apart from numerical factors) when $\gamma=\lambda \mu^{-1}$,  
 provided $|\Psi({\bf p})|^{2}$ 
 only has support for nonrelativistic momenta: then we may 
 set $E\approx M_{i}$, and take it out of 
 the integral in (\ref{9.7})  (the remaining integral$=1$).  
 
 	In the relativistic domain, the most interesting 
feature of (\ref{9.7}) is the dependence of the integrand on $1/E$.  If 
$|\Psi({\bf p})|^{2}$ is fairly well localized in momentum space around ${\bf p}_{0}$, we may set 
$E\approx E_{0}$ and take it out of the integral.  
Thus we see that the collapse rate for a particle 
moving with velocity ${\bf v}_{0}$, as compared to the collapse rate for the particle at rest, is 
smaller by the factor $M_{i}/E_{0}=\sqrt{1-({\bf v}_{0}/c)^{2}}$, i.e., 
it is time dilated\cite{Grassi2}.

\section{Concluding Remarks}\label{sec:Remarks}
  
 	In his famous 1964 paper\cite{Bellthm} John Bell concluded:
 	
 		"In a theory in which parameters are added to quantum mechanics to determine the results of 
individual measurements, without changing the statistical predictions, 
there must be a mechanism whereby the setting of one measurement device 
can influence the reading of another instrument, however remote.  Moreover, 
the signal involved must propagate instantaneously, so that such a theory could not be 
Lorentz invariant."

	In the collapse models discussed here, the additional parameters are the values of the 
fluctuating field w(x) at all spacetime points. The mechanism has two nonlocal aspects.  One is 
the effect of the Probability Rule: it correlates field values at 
spacelike separated points. 
For example, this makes highly likely the survival of only one state
describing a localized clump of particles when the statevector 
starts out in a superposition of
widely separated clump states.  The other is the effect of the direct product: it can 
correlate the states of two different widely separated systems.     
For example, if the statevector starts out as a sum of direct products, 
(an entangled state),  
the collapse suffered by one of the systems brings about the collapse of both to a single product.    
    
	In this paper we have seen that one can construct a Lorentz invariant collapse model.  
We have been led to considerations of tachyons so as to avoid infinite 
spontaneous particle production from the vacuum in lowest order while 
retaining CSL collapse behaviorin the nonrelativistic limit .  In retrospect,   
one might have looked at Bell's second sentence cited above and realized that 
there \emph {is} a natural Lorentz invariant nonlocal structure which 
can propagate instantaneously, and so be led to consider tachyons by another route. 

	The tachyonic structure seems remarkably well matched to the task: 
it is relativistic, it supplies the spacelike four-momenta needed for collapse, it 
eliminates vacuum excitation to lowest order, it allows removal of time-ordering 
so as to eliminate vacuum excitation to all orders without giving  
a vacuous theory (as would its replacement by an ordinary particle, one with timelike four-momentum), 
it forces pair production and annihilation to 
disappear as kinematically untenable thus making the theory finite, it enables 
a model that is local to a high degree of accuracy.  .  

	The present paper shows one way of obtaining a finite relativistic collapse model, so it 
may be considered as an existence proof.    
Elsewhere I shall consider other models that achieve this which  
use time-ordering, e.g., a model  
which employ a relativistically invariant cutoff to 
restrict energy available for vacuum excitation, and which may be applied to particles which interact. 

	 Experiments have suggested  
mass proportional coupling of particles to the fluctuating 
field\cite{PearleSquires,Collett,PRCA} which implies a connection to gravity. Interestingly, 
some string theory models conjoin the 
notions of tachyons and gravity.  It appears that string theorists get rid of the 
tachyons for various good reasons.  But perhaps tachyons 
do have a role to play in fundamental physics and, as Bell suggested\cite{Belllastpaper} 
a connection exists 
between the resolution of the problem of infinities in quantum field theory as 
embodied in the string theory program and the 
resolution of the reality problem.

\acknowledgments

	I would like to dedicate this paper to the memory of my friend Euan Squires and to 
the memory of my friend and student 
Qijia Fu. I am in debt to Renata Grassi for stimulating conversations some 
time ago, and to Qijia Fu, Giancarlo Ghirardi, Adrian Kent, 
Euan Squires and Henry Stapp for helpful 
comments on previous versions of this paper.
 
\appendix
\section{Individual Statevector Collapse}\label{app:A}

	We illustrate here how collapse occurs for individual 
statevectors according to the generalized evolution equation (\ref{3.1}). 
We use the example in section \ref{sec:CSLexample} where, 
instead of the CSL Eq. (\ref{2.5a}), Eq. (\ref{3.1}) yields

\begin{equation}|\psi,T>=\sum_{i} c_{i}|n_{i}>e^{-{1\over 4\lambda}\int_{T_{0}}^{T}dtdt'd{\bf x}d{\bf x}'
	[w({\bf x},t)-2\lambda n_{i}({\bf x})]G(x-x') 
	[w({\bf x}',t)-2\lambda n_{i}({\bf x}')]}\label{A1}\end{equation}
	
\noindent As in section \ref{sec:CSLexample}, set $w({\bf x},t)=2\lambda n_{i}$ 
to show how a $w({\bf x},t)$ 
which makes the exponential in the ith term large ($=1$) can make small the jth exponential:

\begin{equation}e^{-\lambda \int d{\bf x}d{\bf x}'
	[n_{i}({\bf x})- n_{j}({\bf x})]\int_{T_{0}}^{T}dtdt'G({\bf x}-{\bf x}',t-t') 
	[n_{i}({\bf x}')- n_{j}({\bf x}')]}\label{A2}\end{equation}
	
\noindent The time integrals in (\ref{A2}) yield 

	\begin{equation}2(T-T_{0})\int_{0}^{T-T_{0}} d\tau G({\bf x}-{\bf x}',\tau) - 
	2\int_{0}^{T-T_{0}} d\tau \tau G({\bf x}-{\bf x}',\tau)\label{A3}\end{equation}
	
\noindent The first term in (\ref{A3}) grows 
$\sim T$ while the second term eventually approaches a constant 
(assuming the integrals converge as $T\rightarrow \infty$).  Thus, as in 
the example in section 1, the exponential (\ref{A2}) 
approaches zero for $j\not= i$. If $w({\bf x},t)=2\lambda n_{i} + w_{0}({\bf x},t)$, where 
$w_{0}$ almost always fluctuates uniformly about zero, the results are the same.   
In this way the terms in Eq. (\ref{A1}) approach regions 
of disjoint support in $w$-space.  When $w$ lies in the $i$th region, 
$|\psi,T>\rightarrow\sim |n_{i}>$ according to (\ref{A1}), while the integrated probability 
over the $i$th region $\int DwP_{T}(w)\rightarrow |c_{i}|^{2}$.

\section{Fourier Transform Representation}\label{app:B}

   It is well known that the Fourier transform of a Gaussian is a Gaussian:
   
\begin{equation} e^{-{1\over 2\alpha}\sum_{i,j=1}^{N} [w_{i}-a_{i}]G_{ij}[w_{j}-a_{j}]}=
\int{1\over \sqrt {\det G}}\prod_{i=1}^{N} {d\eta _{i}\over \sqrt{2\pi /\alpha}}
e^{-{\alpha\over 2}\sum_{i,j=1}^{N}\eta _{i}G^{-1}_{ij}\eta _{j}}
e^{i\sum_{i=1}^{N}\eta _{i}[w_{i}-a_{i}]}\label{B1}\end{equation}

\noindent where $G^{-1}G=1$.  Applying this to Eq. (\ref{3.1}), with
 
\begin{eqnarray*}\alpha&&\rightarrow 2\gamma\\ 
G_{ij}&&\rightarrow \sqrt{dxdx'}G(x-x')\\  
w_{i}-a_{i}&&\rightarrow \sqrt{dx}[w(x)-2\gamma F(x)]\\
\eta _{i}&&\rightarrow \sqrt{dx}\eta (x)\\  
G^{-1}_{ij}&&\rightarrow G^{-1}_{T_{0},T}(x,x')\\ 
D\eta&&\equiv \Bigl [\det\sqrt{dxdx'}G(x-x')\Bigr ] ^{-{1\over 2}}
\prod_{{\bf x}, t= T_{0}}^{T}
d\eta (x) \sqrt{\gamma /\pi}\end{eqnarray*}

\noindent yields:

\begin{equation}|\psi ,T>={\cal T}\int D\eta e^
{-\gamma\int_{T_{0}}^{T} dxdx'\eta(x)G^{-1}_{T_{0},T}(x,x')\eta(x')}
e^{i\int_{T_{0}}^{T}dx\eta(x)[w(x)-2\gamma F(x)]}|\psi ,{T_{0}}>\label{B2}\end{equation}

\noindent where 

\begin{equation}\int_{T_{0}}^{T}dzG(x-z)G^{-1}_{T_{0},T}(z,x')=\delta (x-x')\label{B3}\end{equation}
	 
\noindent As appears in Eq. (\ref{B3}), $G^{-1}_{T_{0},T}(x,x')$ is the inverse of 
$G(x-x')$ over the interval $(T_{0},T)$.
  
	If $(T_{0},T)$ is $(-\infty,\infty)$, then $G^{-1}$ can be 
written simply in terms of $\tilde G(k)$, the Fourier transform of G:

\begin{equation}G^{-1}(x-x')={1\over (2\pi)^{4}}
\int d^{4}ke^{ik\cdot (x-x')}{1\over \tilde G (k)}\label{B4}\end{equation}

\noindent This simpler form of $G^{-1}$ can be utilized even if $(T_{0},T)$ 
is finite, since (\ref{B1}) may also be used to write  

\begin{equation}|\psi ,T>={\cal T}\int D\eta e^
{-\gamma\int_{-\infty}^{\infty} dxdx'\eta(x)G^{-1}(x-x')\eta(x')}
e^{i\int_{T_{0}}^{T}dx\eta(x)[w(x)-2\gamma F(x)]}|\psi ,{T_{0}}>\label{B5}\end{equation}

\noindent which differs from (\ref{B2}) only in the limits of the 
Gaussian and in $D\eta\sim \prod_{{\bf x}, t=-\infty}^{\infty}d\eta (x)$.  When  $\eta (x)$ 
in (\ref{B5}) is integrated over for $-\infty<t<T_{0}$ and $T<t<\infty$, the result is (\ref{B2}).  
The discrete analog of this can be seen from Eq. (\ref{B1}) if we let the discrete indices
go from $-\infty$ to $\infty$ but let $w_{i}-a_{i}=0$ for $i<1$ and $i>N$.  In 
this case $G^{-1}$ is the inverse of $G$ in the infinite dimensional index space.  
When the integrals over $\eta_{i}$ for $i<1$ and $i>N$ are performed, Eq. (\ref{B1}) is recovered.

	Eqs. (\ref{B4}), (\ref{B5}) are written down on the supposition that $G^{-1}(x-x')$  exists.  
However, we will be applying them when $\tilde G (k)$ vanishes in large enough regions of $k$-space 
so that the integral (\ref{B4}) does not exist.  In this case one may replace $\tilde G (k)$ by
a function $\tilde G_{\epsilon}(k)$ which approaches $\tilde G (k)$ as 
$\epsilon\rightarrow 0$ and which does exist, and take the limit 
$\epsilon\rightarrow 0$ after all calculations are completed. 
Alternatively, one may use the form (\ref{B2}) with $G^{-1}_{T_{0},T}(x,x')$ 
which generally does exist.   
In practice one doesn't have to worry about these niceties since 
perturbation expressions 
involve $G$ and not $G^{-1}$.  Thus formal manipulations suffice to give sensible answers.
  
	The Fourier transform representation (\ref{3.5}) of the density matrix may be derived 
by applying (\ref{B1}) to the density matrix expression (\ref{3.3}). More easily, 
we may use the statevector 
representation (\ref{B1}) in $\rho (T) =\int Dw|\psi ,T><\psi ,T|$: 
integration over $w(x)$ for each $x$ 
gives rise to a term $\sim \prod_{x} \delta (\eta(x)-\eta'(x))$, and integration over $\eta'(x)$ 
for each $x$ results in (\ref{3.5}).

\section{Lorentz Invariance}\label{app:C}
 
 	We wish to show that the statevector evolution (\ref{3.1}) 
 		
\begin{equation}|\psi ,\sigma>={\cal T} e^{-{1\over 4\gamma}\int_{\sigma_{0}}^{\sigma}dxdx'
   			[w(x)-2\gamma F(x)]G(x-x')[w(x')-2\gamma F(x')]}
   			|\psi ,\sigma_{0}>\label{C1}\end{equation}
   			
 \noindent and the probability rule (\ref{2.2}) describe a Lorentz invariant theory.  
 For RCSL$_{1}$, where $G(x-x')=\delta (x-x')$, it has been shown that 
 the theory is Lorentz invariant\cite{GGP}.  However, RCSL$_{1}$ possesses the 
 Markovian property that two successive transformations from $\sigma_{0}$ to $\sigma_{1}$ and 
 from $\sigma_{1}$ to $\sigma_{2}$ are equivalent to one transformation from 
 $\sigma_{0}$ to $\sigma_{2}$, and this is a property not possessed in the more general case.  
 
 	We first discuss Lorentz invariance from the passive point of view, 
and show that the evolution of 
$|\psi ,\sigma_{0}>$ to $|\psi ,\sigma>$ is described identically in all Lorentz frames. 
Consider a Lorentz frame with coordinates $\bar x =Lx$.  Since $\bar F (\bar x )=F(x)$, 
$\bar w (\bar x )=w(x)$, $d\bar x=dx$, and $G(\bar x-\bar x')=G(x-x')$, 
we see that (\ref{C1}) 
is equal to the same expression when unbarred quantities are replaced by barred quantities 
(the hypersurfaces $\sigma_{0}$, $\sigma$ are not changed although, of course, their 
expressions in terms of $x$ and $\bar x$ differ).  Finally, the probability density 
$<\psi ,\sigma|\psi ,\sigma>$ is invariant, as is the functional integration element 
$D\bar w(\bar x)=Dw(x)$.

	Next we discuss Lorentz invariance from te active point of view and show that, if $U$ is the
unitary transformation corresponding to $L^{-1}$, then 

\begin{equation}U|\psi ,\sigma>={\cal T} e^{-{1\over 4\gamma}\int_{\bar \sigma_{0}}^{\bar \sigma}dxdx'
   			[w'(x)-2\gamma F(x)]G(x-x')[w'(x')-2\gamma F(x')]}U|\psi ,\sigma_{0}>\label{C2}\end{equation}

	Operating upon (\ref{C1}) with $U$ replaces $|\psi ,\sigma>$ by $U|\psi ,\sigma>$, 
$|\psi ,\sigma_{0}>$ by $U|\psi ,\sigma_{0}>$, and $F(x)$ by $UF(x)U^{-1}=\bar F (x)$.  Next, 
change the {\it labels} from $x$ to $\bar x$ and $x'$ to $\bar x'$.  
This is {\it not} a Lorentz transformation, 
but a consistent notation requires changing the integration limits from  ($\sigma_{0},\sigma)$ to 
($\bar \sigma_{0},\bar \sigma)$.  (For example, if ($\sigma_{0},\sigma)=(T_{0}, T)$ 
are the integration limits of $t$, and 
$L$ is the time translation $\bar t =t-\tau$, then a 
relabelling $t\rightarrow \bar t$ gives $(T_{0}, T)$ as the 
integration limits for $\bar t$, which are different hyperplanes than $t=(T_{0}, T)$.)  
Now we Lorentz transform back to the unbarred frame, using $\bar F (\bar x )=F(x)$, 
$d\bar x=dx$ and $G(\bar x-\bar x')=G(x-x')$, obtaining (\ref{C2}), with $w'(x)\equiv w(Lx)$.  
Although $w'(x)$ is not equal to $w(x)$, that does not matter since $w(x)$ is an arbitrary 
classical field, so the sets $\{ w(x)\}$, $\{ w'(x)\}$ are equivalent.  
Indeed, this is all that is necessary for the probability rule to be invariant.

\section{Vacuum Excitation to Order $\gamma$ in RCSL$_{1}$.}\label{app:D}
\bigskip

 	We calculate the contribution of the diagrams 2a,b,c  
to the density matrix at ``large" time $T/2$ (i.e., $T>>\hbar/\mu c^{2}\approx 10^{-15}$sec). 
With the 
initial state as the vacuum state, Eq. (\ref{4.1})  (with the help 
of (\ref{4.4})) gives to lowest order 
 	 
\begin{equation}\rho(T/2)=|0><0|-\gamma\int_{-T/2}^{T/2} dxdx'G(x-x'){\cal T}
[\phi(x)[\phi(x'),|0><0|]]\label{D1}\end{equation} 
 
\noindent Using Wick's theorem, and remembering that ${\cal T}$ means operators 
to the right of $|0><0|$ are time-reversed, yields 

\begin{eqnarray}\rho(T/2)&&=|0><0|
-\gamma\int_{-T/2}^{T/2} dxdx'G(x-x')\nonumber\\
&&\qquad\cdot\Biggl[{1\over (2\pi )^{3}}\int dke^{ik\cdot (x-x')}\delta (k^{2}-\mu^{2})|0><0|
 -2\phi_{-}(x)|0><0|\phi_{+}(x')\Biggr]\label{D2}\end{eqnarray}

\noindent where $\phi_{+}(x)$, $\phi_{-}(x)$ are 
respectively the positive and negative frequency parts of $\phi$.
The first term in the bracket of (\ref{D2}) corresponds to the sum of the diagrams 
2a,b, with $\delta (k^{2}-\mu^{2})$ coming from the sum of the $\phi$ propagator 
and its complex conjugate.  The last term in the bracket 
corresponds to 2c.  When we perform the spacetime integrals, we obtain 

\begin{equation}
\rho(T/2)=|0><0|-\gamma \tilde G(\mu^{2})T\int {d{\bf k}\over \omega ({\bf k})}
\Biggl[{V\over (2\pi )^{3}}|0><0|-a^{\dagger}({\bf k})|0><0|a({\bf k})\Biggr]\label{D3}
\end{equation}

\noindent (we have used 
$2\pi\delta (\omega =0)=\int_{-T/2}^{T/2}dt\exp i0t=T$, 
and similarly $(2\pi)^{3}\delta ({\bf k}=0)=V$). 

	The mean energy 
$\bar H (T/2)\equiv$Tr$\int d {\bf k}\omega ({\bf k})a^{\dagger}({\bf k})a({\bf k})\rho (T/2)$ is 
found using (D3) to be 

	\begin{equation}\bar H (T/2)=\gamma \tilde G(\mu^{2}) TV {1\over (2\pi)^{3}}
	\int d{\bf k}\label{D4}
	\end{equation}
	
\noindent It is apparent from Eq. (\ref{D4}) that 
the energy/sec-vol produced out of the vacuum is infinite 
(as in RCSL$_{1}$, where $\tilde G=1$) unless $\tilde G(\mu^{2})=0$. 
Then, according to Eq. (\ref{D3}), the 
vacuum remains the vacuum in order $\gamma$.

\section{Nonlocal Transport In RCSL}\label{app:E}
\subsection{No-Time-Ordering Expressed As Time-Ordering}\label{app:E1}

 	We may express the not-time-ordered 
 evolution as a time-ordered evolution.  The following is a 
 well--known identity\cite{Liegroups}:
 	
\begin{equation}{d\over dt}e^{A(t)}=\left\{\int_{0}^{1}d\alpha e^{\alpha [A(t)\otimes 1-1\otimes A(t)]}
{dA(t)\over dt}\right\} e^{A(t)}\label{E1}
\end{equation}

\noindent Its integral

\begin{equation}e^{A(T)}={\cal T}e^
{\int_{0}^{T} dt\int_{0}^{1}d\alpha e^{\alpha [A(t)\otimes 1-1\otimes A(t)]}{dA(t)\over dt}}
\label{E2}\end{equation}

\noindent (assuming $A(T_{0})=0$) is the relation we need.  Put

\begin{equation}A(t)=-i2\gamma(2M)\int_{T_{0}}^{t}dx'\eta(x')
:\negthinspace\negthinspace\psi_{t}^{2}(x')\negthinspace\negthinspace:\label{E3}\end{equation}

\noindent into (\ref{E2}).  (By  $\psi_{t}(x')$ we mean $\psi(x')$ for all purposes except 
time-ordering, in which case it should be regarded as a function of $t$: this is 
required by Eqs. (\ref{E2},\ref{E3}), since $A(t)$ must be regarded as a function of $t$ 
for time-ordering.)  Thus we obtain the unitary operator part of the integrands 
in Eqs. (\ref{7.3},\ref{7.4})  
which has no time ordering, written in a time-ordered way:

\begin{equation}e^{-i4\gamma M\int_{T_{0}}^{T}dx\eta (x) 
:\negthinspace\negthinspace\psi_{t}^{2}(x')\negthinspace\negthinspace:}
={\cal T}e^{-i\int_{T_{0}}^{T}dt H_{eff}(t)}\label{E4}\end{equation}

\begin{mathletters}\label{E5}
\begin{eqnarray}\int_{T_{0}}^{T}dt&& H_{eff}(t)\equiv 4\gamma M\int_{0}^{1}d\alpha
\int_{T_{0}}^{T}dx\eta (x)e^{-i4\gamma M\alpha \int_{T_{0}}^{t}dx'\eta (x') 
[:\psi_{t}^{2}(x'):\otimes 1-
1\otimes :\psi_{t}^{2}(x'):]}
:\negthinspace\negthinspace\psi^{2}(x)\negthinspace\negthinspace:\label{E5a}\\
&&=4\gamma M\int_{T_{0}}^{T}\eta (x)
:\negthinspace\negthinspace\psi^{2}(x)\negthinspace\negthinspace:\nonumber\\
&&\quad+(4\gamma M)^{2}\int_{T_{0}}^{T}dx\int_{T_{0}}^{t}dx'
\eta (x)\eta (x')i[\psi(x),\psi (x')]\{\psi (x),\psi_{t} (x')\}_{+}\nonumber\\
&&\quad+{(4\gamma M)^{3}\over 3}\int_{T_{0}}^{T}dx\int_{T_{0}}^{t}dx'dx''
\eta (x)\eta (x')\eta (x'')i[\psi(x),\psi (x')]\cdot\nonumber\\
&&\qquad\Bigl[i[\psi(x),\psi (x'')]\{\psi_{t}(x'),\psi_{t} (x'')\}_{+}
+i[\psi(x'),\psi (x'')]\{\psi(x),\psi_{t} (x'')\}_{+}\Bigr]+...\label{E5b}\end{eqnarray}
\end{mathletters}

	All terms in Eq. (\ref{E5b}) are quadratic in the operator $\psi$ (the 
commutators are c--numbers).  The first term 
is local.  The 
remaining terms act like a nonlocal potential: 
a particle may be transported from $x$ to $x'$. The second term, of order 
$\gamma^{2}$, vanishes for spacelike $x-x'$ so, although it is nonlocal, it 
describes timelike transport. However, the transport 
may be in the backwards time direction and, when iterated, a sequence of timelike forward and 
backward motion     
may allow spacelike transport. The third and higher order terms, 
may also allow spacelike transport, due to a 
succession of commutators which have support in the forward and backward light cones. 
However, this discussion does not take into account 
the nature of the field $\eta (x)$ which appears in these expressions.  Whether spacelike 
transport actually occurs in any term requires further investigation. 

\subsection{Particle Propagation}\label{app:E2}

  Here we calculate to lowest order the propagation of a particle 
initially located near the origin (say $|{\bf x}|<\sigma$) and with a fairly well-defined 
momentum, corresponding to diagram 4c, for a finite time $T$.    
We take a pragmatic approach to the problems of defining a position 
operator and a localized wavefunction in a relativistic theory. We use the 
usual (nonrelativistic) position operator ${\bf X}$ with eigenstates $|{\bf x}>$, choose 
the wavepacket width $<<M^{-1}$, and regard a wavefunction's exponential decay 
outside the light cone (with characteristic length $M^{-1}$) as normal, i.e., as not indicative of 
superluminal transport. 

	The desired probability is given by Eq. (\ref{7.5}) as: 
	
\begin{equation}\sim \gamma M^{2}\int_{0}^{T}dxdx'G(x-x')
<{\bf z},T|\psi_{-}(x')\psi_{+}(x')|\Phi><\Phi|\psi_{-}(x)\psi_{+}(x)|{\bf z}, T>
\label{E6}\end{equation}

\noindent In this and what follows we shall drop purely numerical factors.  Here, 
$<{\bf k}|\Phi>$ is the initial wavefunction in momentum space and $|{\bf z},T>$ is 
an eigenstate of 
the interaction picture position operator ${\bf X}(T)$
($<{\bf z},T|{\bf k}>=\exp i({\bf k}\cdot{\bf z}-ET)$). After some 
integrations we obtain

\begin{eqnarray}&&\sim \gamma M^{2}\int{d{\bf p}d{\bf k}d{\bf k}'e^{i({\bf k}-{\bf k}')
\cdot{\bf z}
-i(E_{{\bf k}+{\bf p}}-E_{{\bf k}'+{\bf p}})T}
\over \epsilon_{\bf p}\sqrt{E_{{\bf k}}
E_{{\bf k}+{\bf p}}E_{{\bf k}'}E_{{\bf k}'+{\bf p}}}}<{\bf k}|\Phi><\Phi|{\bf k}'>\nonumber\\ 
&&\qquad\qquad\qquad\qquad\cdot\sum_{s=-1}^{1}\int_{0}^{T}\int_{0}^{T}dtdt'
e^{i[E_{{\bf k}+{\bf p}}-E_{{\bf k}}-s\epsilon_{\bf p}]t}
e^{-i[E_{{\bf k}'+{\bf p}}-E_{{\bf k}'}-s\epsilon_{\bf p}]t'}\label{E7}\end{eqnarray}

\noindent where $\epsilon_{\bf p}\equiv\sqrt{{\bf p}^{2}-\mu^{2}}$, 
$E_{\bf k}\equiv\sqrt{{\bf k}^{2}+M^{2}}$.

	We may use one of the time integrals to give energy conservation: 
	
\begin{equation}\int_{0}^{T}dte^{i\Delta E t}={\sin(\Delta E T)\over\Delta E}
+2i{\sin^{2}(\Delta E T/2)\over\Delta E}
\approx{\pi\over 2}\delta(\Delta E)+2i{\cal P}
{1\over \Delta E}\approx{\pi\over 2}\delta(\Delta E)\label{E8}\end{equation}

\noindent The term ${\cal P}1/\Delta E$ is dropped in  
(\ref{E8}) as it is a well known artifact of the abrupt turning on and off 
of the interaction and can be eliminated by more careful attention to that detail. 
(\ref{E8}) is a good approximation provided the rest of the integrand does not oscillate in energy 
with period greater than $\approx\hbar/T\approx 6\cdot10^{-16}$eVsec$/T$.  Even 
for a relatively short time such as $T\approx 10^{-14}$sec this gives energy conservation to 
better than $.1$eV (which improves with increasing T). (\ref{E8}) can be used to integrate over 
$\epsilon_{\bf p}$ because the dependence of the rest of the integrand is upon 
${\bf p}=\sqrt{\epsilon ^{2}+\mu ^{2}}$ which varies slowly over e.g., $\Delta \epsilon=.1$eV.

	We shall also assume that $<{\bf k}|\Phi>\equiv\tilde\Phi ({\bf k}-{\bf k}_{0})$ is 
peaked at ${\bf k}_{0}$.  The kinematics of absorption/emission of a tachyon 
by a particle initially moving with momentum ${\bf k}$ can be shown to yield 
the tachyon momentum magnitude (which is the magnitude of the increase in particle momentum) 

\begin{equation}|{\bf p}({\bf k},\hat{\bf p})|=
{E_{\bf k}\mu\over \sqrt{({\bf k}\times\hat{\bf p})^{2}+M^{2}}}\label{E9}\end{equation}

\noindent to o$(\mu/M)$ or better.  Since $|{\bf p}|<<E_{\bf k}$,  
and if $\tilde\Phi$'s peak is narrow,  
we can expand $E_{{\bf k}+{\bf p}}$ to second order in ${\bf p}$ and 
${\bf\kappa}\equiv {\bf k}-{\bf k_{0}}$.  (\ref{E7}) then becomes (with replacement of 
${\bf k_{0}}$ by ${\bf k}$, and $E_{\bf k}$ by $E$):

\begin{mathletters}\label{E10}
\begin{eqnarray}&&\sim {\gamma M^{2}\over E^{2}}\int|{\bf p}({\bf k},\hat{\bf p})|
d\Omega_{{\bf p}}d{\bf\kappa} d{\bf\kappa}'
e^{i({\bf\kappa}-{\bf\kappa}')\cdot[{\bf z}-({\bf v}+{\bf w})]T}
\tilde\Phi({\bf\kappa})\tilde\Phi^{*}({\bf\kappa}') 
\int_{0}^{T}dte^{i({\bf\kappa}-{\bf\kappa}')\cdot{\bf w}t}\label{E10a}\\
&&\sim {\gamma M^{2}\over E^{2}}\int|{\bf p}({\bf k},\hat{\bf p})|
d\Omega_{{\bf p}}\int_{0}^{T}dt|\Phi({\bf z}-{\bf v}T-{\bf w}(T-t))|^{2}\label{E10b}\\
&&\approx{\gamma M^{2}\over E^{2}}\int|{\bf p}({\bf k},\hat{\bf p})|
d\Omega_{{\bf p}}\int_{0}^{T}dt\delta[{\bf z}-{\bf v}T-{\bf w}(T-t)]\label{E10c}
\end{eqnarray}
\end{mathletters}

\noindent  where

\begin{equation}{\bf v}\equiv{{\bf k}\over E}\qquad \hbox{and}\qquad 
{\bf w}\equiv {{\bf p}\over E}-{{\bf p}\cdot{\bf k}{\bf k}\over E^{3}}\label{E11}\end{equation}

\noindent are the initial velocity of the particle and its velocity gain respectively. 
In the integrand of (\ref{E10a}), for simplicity we have 
omitted the factor $\exp-i({\bf\kappa}^{2}-{\bf\kappa}'^{2})T$  
responsible for the usual spread of the wavepacket: we assume that $\sigma$ is 
large enough ($\sigma>a)$ so that this spread is less than the 
spread described by (\ref{E10}).  On the other hand, we assume 
that $\sigma<<$ the spread described by (\ref{E10}), so that $\Phi^{2}$ can be well 
approximated by the delta function in (\ref{E10c}).  

	In the nonrelativistic limit, ${\bf p}=\mu c\hat{\bf p}$ and ${\bf w}=(\mu c/M)\hat{\bf p}$
by Eqs. (\ref{E9}), (\ref{E11}).  
With the substitution $\hat{\bf p}(T-t)\equiv{\bf r}$, the integral over ${\bf r}$ in (\ref{E10c}) is 
easily performed with the result:

\begin{equation}\sim{\gamma M\over|{\bf z}-{\bf v}T|^{2}}
\Theta[{\mu\over M}cT-|{\bf z}-{\bf v}T|]\label{E12}\end{equation}

\noindent ($\Theta$ is the step function.)  This is precisely the same distribution as would be 
obtained from a collection of classical particles, all of which start at $t=0$ 
from ${\bf z}=0$ with velocity ${\bf v}$, each of which impulsively acquires 
momentum of magnitude $\mu cT/ M$ in a random direction at an arbitrary time between 0 and $T$.

	This result is used in section \ref{sec:ParticleMotion} to explain how 
a succession of such evolutions in sufficiently high order (and so with very small probability) will 
eventually go outside of the lightcone.  We note here that the above calculation does not 
take into account the free backwards-in-time evolution because it calculates 
the wavefunction in the basis $|{\bf z},T>$ and not in the basis $|{\bf z},0>$.  The latter 
is the appropriate basis to use when iterating diagrams as it introduces 
no extra time evolution beyond what the expansion provides (i.e., it 
includes the free backwards-in-time evolution).  
The former is to be used to find the probability of the particle's position ($|{\bf z},T>$ is 
an eigenstate of the interaction picture position operator)
only at the end of the calculation of the contribution 
of each order, where its use correctly eliminates the final free backwards-in-time evolution.

	With relativistic ${\bf k}$, first consider ${\bf k}$ parallel to 
$\hat{\bf u}\equiv[{\bf z}-{\bf v}T]/ |{\bf z}-{\bf v}T|$.  The delta function in (\ref{E10c}) 
can only vanish for ${\bf w}\parallel\hat{\bf u}$ which, by (\ref{E11}), implies that 
${\bf p}\parallel\hat{\bf u}$.  Then $|{\bf p}|=E\mu/M$ by (\ref{E9}), 
and ${\bf w}=\hat{\bf p}\mu M/E^{2}$ by (\ref{E11}). 
The integral in (\ref{E10c}) can then be performed with the result:

\begin{equation}\sim{\gamma E\over|{\bf z}-{\bf v}T|^{2}}
\Theta[{\mu M\over E^{2}}cT-|{\bf z}-{\bf v}T|]\label{E13}\end{equation}

	With ${\bf k}\perp\hat{\bf u}$, then ${\bf k}\cdot{\bf w}=0$ 
(since ${\bf w}\parallel\hat{\bf u}$) which implies 
${\bf k}\cdot{\bf p}=0$ by (\ref{E11}).  Then $|{\bf p}|=\mu$ by (\ref{E9}) 
and ${\bf w}=\hat{\bf p}(\mu/E)$ by (\ref{E11}). (\ref{E10c}) then yields: 

\begin{equation}\sim{\gamma M^{2}\over|{\bf z}-{\bf v}T|^{2}}
\Theta[{\mu \over E}cT-|{\bf z}-{\bf v}T|]\label{E14}\end{equation}

\noindent In the nonrelativistic limit, Eqs. (\ref{E13}), (\ref{E14}) agree with the 
result (\ref{E12}). They describe (to accuracy o$(\mu /M)$) the diminution in 
the wavepacket spread which is the effect of a Lorentz 
transformation on the zero velocity spread..  

	As an interesting application of the 
relativistic results, consider the spreading 
of a wavepacket when one one starts with a particle at rest and in each order 
has all ${\bf p}$'s  parallel (i.e., one does does not perform the integration over the 
${\bf p}$'s). This accelerates the particle to an arbitrarily large energy 
but, as mentioned in section \ref{sec:ParticleMotion}, the result (see (\ref{E15}))  
is that the wavefunction does not go outside 
the lightcone. I shall not go into the details 
of how the successive iterations (the higher order terms) 
in the perturbation expansion work: in (\ref{E6}) one must calculate the matrix element between
$<{\bf z},0|...|{\bf z}', 0>$ in order to obtain the 
wavefunction appropriate for the next iteration, etc.  
The momentum gained by the particle 
with each successive iteration, given by (\ref{E9}), is $\Delta k=E\mu/M$.   
The increase in spreading in the direction of the accumulating 
momentum, given by (\ref{E13}), is $\Delta  L=\Delta k M^{2}T/E^{3}$.  Integrating over $k$ from 
0 to $\infty$ gives the upper limit on the accumulated wavefunction spread for any 
number of iterations:

\begin{equation}L<cT\int_{0}^{\infty}dk{M^{2}\over E^{3}}=cT\label{E15}\end{equation}

\noindent  
 	   
\section{ Relativistic Energy Production to Order $\gamma$.}\label{app:F}
\bigskip

	We calculate here $\bar H(T/2)\equiv\hbox{Tr}H\rho(T/2)$, for an initial $n$  
particle state  $\rho(-T/2)=|\Psi_{0}><\Psi_{0}|$. By Eq. (\ref{7.5}), 

\begin{eqnarray}
\bar H(T/2)-\bar H(-T/2)&&=-2\gamma(2M)^{2}\int_{-{T\over 2}}^{T\over 2}dxdx'G(x-x')\nonumber\\
&&\qquad\qquad\cdot<\Psi_{0}|[\psi_{-}(x)\psi_{+}(x)
[\psi_{-}(x')\psi_{+}(x'), H]]|\Psi_{0}>
\label{F1}\end{eqnarray}

\noindent Replacement of $[\psi_{-}(x')\psi_{+}(x'), H]$ by 
$i\partial\psi_{-}(x')\psi_{+}(x')/\partial t'$ followed by integration by parts over $t'$ (the 
surface terms are exponentially small for large $T$, and may be neglected),  
and evaluation of the remaining commutator results in

\begin{eqnarray}\bar H(T/2)-\bar H(-T/2)&&=-4i\gamma (2M)^{2}
\int_{-{T\over 2}}^{T\over 2}dxdx'{\partial\over \partial t} G(x-x')\nonumber\\
&&\qquad\qquad\cdot<0|\psi_{+}(x)\psi_{-}(x')|0>
<\Psi_{0}|\psi_{-}(x)\psi_{+}(x'))|\Psi_{0}>
\label{F2}\end{eqnarray}

\noindent Next, the integral over $x$, $x'$  gives 

\begin{eqnarray}\bar H(T/2)&&-\bar H(-T/2)={1\over \pi^{2}}\gamma (2M)^{2}
\int dk{d{\bf p}\over 2E}{d{\bf p}_{1}\over \sqrt{2E_{1}}}{d{\bf p}_{2}\over\sqrt{2E_{2}}}\nonumber\\
&&\cdot k^{0}\delta (k^{2}+\mu^{2})\delta (k-p+p_{1})\delta (k-p+p_{2})
<\Psi_{0}|a^{\dagger}({\bf p}_{1})a({\bf p}_{2})|\Psi_{0}>\label{F3}
\end{eqnarray}

\noindent We may set $\delta (p_{1}-p_{2})=\delta ({\bf p}_{1}-{\bf p}_{2})T/2\pi$ and perform the 
integral over $p_{2}$:

\begin{eqnarray}\bar H(T/2)-\bar H(-T/2)=&&{1\over 2\pi^{3}}\gamma T (2M)^{2}
\int {d{\bf p}_{1}\over 2E_{1}}<\Psi_{0}|a^{\dagger}({\bf p}_{1})a({\bf p}_{1})|\Psi_{0}>\nonumber\\
&&\ \cdot \int dkk^{\nu}\delta (k^{2}+\mu^{2})
\int {d{\bf p}\over 2E}\delta (k-p+p_{1})|_{\nu =0}
\label{F4}\end{eqnarray}

	The integral expression in the last line of (\ref{F4}) 
is a four--vector function of $p_{1}$, so it is equal 
to $p_{1}^{\nu}C$, where $C$ is a constant.  The zeroth component of this four--vector is 
$E_{1}C$, so the integral over ${\bf p}_{1}$ yields the particle number operator, whose matrix 
element is $n$.  Thus (\ref{F4}) becomes

\begin{equation}\bar H(T/2)-\bar H(-T/2)={1\over \pi^{3}}n\gamma T M^{2}C\label{F5}\end{equation}

\noindent C is most easily evaluated in the Lorentz frame in which $p_{1}=(M,{\bf 0})$:

\begin{equation}MC=\int {d{\bf k}\over 2\sqrt{{\bf k}^{2}+\mu^{2}}}dk^{0}k^{0}
\delta (k^{2}+\mu^{2})\delta (k^{0}-\sqrt{{\bf k}^{2}+M^{2}}+M)=
M{\pi\over 2}\biggl({\mu\over M}\biggr)^{3}\sqrt{1+\biggl({\mu\over 2M}\biggr)^{2}}\label{F6}\end{equation}

\noindent Eqs. (\ref{F5}) and (\ref{F6}) result in Eq. (\ref{9.2}).

\section{Relativistic Collapse Rate to Order $\gamma$.}\label{app:G}

	Consider a particle in the initial state $\alpha|L>+\beta|R>$, where $|L>$ is a reasonably 
well localized state around ${\bf x}={\bf L}$ and, for simplicity, $|R>$,  
is the same state translated by a 
large distance to ${\bf R}$, with $|{\bf L}-{\bf R}|>>cT$.  Since  
$\psi_{-}({\bf x},0)\psi_{+}({\bf x},0)|L>$ is a state still localized in the region around 
${\bf L}$ (the operator is a kind of particle number density operator smeared over the 
distance $M^{-1}$) then, even when t is not $0$, 

\[< R|\psi_{-}(x)\psi_{+}(x)|L>\approx 0, \qquad 
< R|\psi_{-}(x)\psi_{+}(x)\psi_{-}(x')\psi_{+}(x')|L>\approx 0\] 

\noindent by our hypothesis of large separation of the states. This helps simplify the 
density matrix expression (\ref{7.5}):

\begin{eqnarray}&&<L|\rho (T/2)|R>=\alpha\beta^{*}\Bigl\{ 1-2\gamma(2M)^{2}
\int dxdx'G(x-x')\nonumber\\
&&\quad \cdot[ <L|\psi_{-}(x)\psi_{+}(x)\psi_{-}(x')\psi_{+}(x')|L> +
<R|\psi_{-}(x)\psi_{+}(x)\psi_{-}(x')\psi_{+}(x')|R>\nonumber\\
&&\qquad\qquad -2
<L|\psi_{-}(x)\psi_{+}(x)|L><R|\psi_{-}(x')
\psi_{+}(x')|R>] \Bigr\}\label{G1}\end{eqnarray}

\noindent The first two terms in the square bracket in (\ref{G1}) correspond to the 
Feynman diagrams in Fig. 4a, the last term to Fig. 4c.  

  The last term in the bracket of (\ref{G1}) has 
the form $f({\bf x}-{\bf L},t)f({\bf x'}-{\bf R},t')$ 
where $f({\bf z}, 0)$ is fairly well localized, around ${\bf z}=0$. $G$ decreases
sufficiently rapidly (see Eqs. (\ref{5.4})) so that the contribution of the last term vanishes 
as $|{\bf L}-{\bf R}|\rightarrow\infty$.  Because of translation invariance, the first two matrix 
elements in the bracket of (\ref{G1}) are identical.  After commuting the two middle operators in the 
matrix element (the commutator gives no contribution 
because its support vanishes where $G$'s does not), we obtain

\begin{eqnarray}<L|\rho (T/2)|R>&&=\alpha\beta^{*}
\Bigl\{ 1-4\gamma(2M)^{2}\int dxdx'G(x-x')\nonumber\\
 &&\qquad\qquad\qquad\qquad
 \cdot <0|\psi_{+}(x)\psi_{-}(x')|0>
 <L|\psi_{-}(x)\psi_{+}(x')|L>\Bigr\}\label{G2}\end{eqnarray}

	We next perform the integrals over $x$ and $x'$:
	
\begin{eqnarray}&&<L|\rho (T/2)|R>=\alpha\beta^{*}\Bigl\{ 1-{1\over\pi^{2}}\gamma (2M)^{2}
\int dk{d{\bf p}\over 2E}{d{\bf p_{1}}\over \sqrt{2E_{1}}}{d{\bf p_{2}}\over \sqrt{2E_{2}}}\nonumber\\
 &&\qquad\cdot \delta (k^{2}+\mu^{2})\delta (k-p+p_{1})\delta (p_{1}-p_{2})
<L|a^{\dagger}({\bf p_{2}})|0><0|a({\bf p_{1}})|L>\Bigr\}\label{G3}\end{eqnarray}
 
\noindent We shall write the normalized wavefuction in momentum space as 
$<0|a({\bf p_{1}})|L>\equiv\Psi({\bf p_{1}})$.  Since 
$\delta (p_{1}-p_{2})=\delta ({\bf p_{1}}-{\bf p_{2}})T/2\pi$, we get 

\begin{equation}<L|\rho (T/2)|R>=\alpha\beta^{*}\Bigl\{ 1-{1\over2\pi^{3}}\gamma T(2M)^{2}
\negthinspace\negthinspace\int{d{\bf p_{1}}\over {2E_{1}}}|\Psi({\bf p_{1}})|^{2}
\negthinspace\negthinspace\int dk\delta (k^{2}+\mu^{2})
\negthinspace\negthinspace\int{d{\bf p}\over 2E}\delta (k-p+p_{1})\Bigr\}\label{G4}\end{equation}

\noindent The last two integrals in (\ref{G4}) are a Lorentz scalar function of $p_{1}^{2}=M^{2}$, 
i.e., they do not depend upon $p_{1}^{\mu}$ at all.  The integrals are most easily performed in 
the frame in which $p_{1}=(M,{\bf 0})$:

\begin{equation} \int dk\delta (k^{2}+\mu^{2})\int{d{\bf p}\over 2E}\delta (k-p+p_{1})=
\pi{\mu\over M}\sqrt{1+\biggl({\mu\over 2M}\biggr)^{2}}\label{G5}\end{equation}

\noindent Thus we obtain the off-diagonal elements of the density matrix as

\begin{equation}<L|\rho (T/2)|R>=\alpha\beta^{*}\Bigl\{ 1-{2\over\pi^{2}}\gamma T\mu M
\sqrt{1+\biggl({\mu\over 2M}\biggr)^{2}}
\int{d{\bf p}\over {2E}}|\Psi({\bf p})|^{2}\Bigr\}\label{G6}\end{equation}

	While some of the depletion of the initial states $|L>$ and $|R>$ is due to their excitation 
and not collapse, Eq. (\ref{G6}) still serves to give a good measure of the collapse rate.

\begin{figure}

\caption{Diagram parts: 	
a) The vertex  $2\gamma\eta\phi$.
b) The vertex $g(2M)\phi:\negthinspace\negthinspace\psi^{2}\negthinspace\negthinspace:$.
 c) The "noise propagator" G attached to two $\eta$ vertices.}\label{fig1} 
 
\end{figure}

\begin{figure}

\caption{ Diagrams involved in particle production from the vacuum, for RCSL$_{1}$.	
a)--c) $\phi$ particle production in order $\gamma$.
d) Particle pair production in order $\gamma g^{2}$. 
e) $\phi$ pair production in order $\gamma g^{6}$.}\label{fig2}

\end{figure}

\begin{figure}

\caption{Diagrams responsible for collapse in (lowest) order $\gamma g^{2}$, 
for RCSL$_{1}$.}\label{fig3}  

\end{figure}

\begin{figure}

\caption{Diagrams responsible for collapse in (lowest) order $\gamma$, 
for RCSL$_{2}$ and RCSL.}\label{fig4}    

\end{figure}

\begin{figure}

\caption{Pair production diagram in order $\gamma^{2}$.}\label{fig5} 
	
\end{figure}

\end{document}